\documentclass{ecai} 

\usepackage{ragged2e} 

\makeatletter
\long\def\@makecaption#1#2{%
  \vskip\abovecaptionskip
  \sbox\@tempboxa{{\captsize\bfseries #1.\quad #2}}%
  \ifdim \wd\@tempboxa >\hsize
    {\captsize\justifying\noindent{\bfseries #1.}\quad #2\par}%
  \else
    \global \@minipagefalse
    \hbox to\hsize{\hfil\box\@tempboxa\hfil}%
  \fi
  \vskip\belowcaptionskip}
\makeatother

\usepackage{latexsym}
\usepackage{amssymb}
\usepackage{amsmath}
\usepackage{amsthm}
\usepackage{booktabs}
\usepackage{enumitem}
\usepackage{graphicx}
\usepackage{color}
\usepackage{url}            
\usepackage{amsfonts}       
\usepackage{nicefrac}       
\usepackage{bookmark}
\usepackage{microtype}      
\usepackage{subcaption} 
\usepackage{algpseudocode}
\usepackage{times}
\usepackage{soul}
\usepackage{url}
\usepackage{algorithm}
\usepackage[utf8]{inputenc} 
\usepackage[T1]{fontenc}    
\newtheorem{remark}{Remark}

\DeclareUnicodeCharacter{03B1}{\ensuremath{\alpha}}

\begin{document}


\begin{frontmatter}


\paperid{4998} 


\title{Fusion-PSRO: Nash Policy Fusion for \\ Policy Space Response Oracles}



\author[A]{\fnms{Jiesong}~\snm{Lian}\footnote{Equal contribution.}}
\author[B]{\fnms{Yucong}~\snm{Huang}\footnotemark[1]}
\author[C]{\fnms{Chengdong}~\snm{Ma}}
\author[C]{\fnms{Mingzhi}~\snm{Wang}}
\author[D]{\fnms{Ying}~\snm{Wen}\thanks{Co-corresponding Author. Email: ying.wen@sjtu.edu.cn.}}
\author[A,F]{\fnms{Long}~\snm{Hu}}
\author[A,F]{\fnms{Yixue}~\snm{Hao}\thanks{Co-corresponding Author. Email: yixuehao@hust.edu.cn.}}

\address[A]{Huazhong University of Science \& Technology, Wuhan, China}
\address[B]{School of Software and Microelectronics, Peking University, Beijing, China}
\address[C]{Institute for Artificial Intelligence, Peking University, Beijing, China}
\address[D]{Shanghai Jiao Tong University, Shanghai, China}
\address[F]{Guangdong HUST Industrial Technology Research Institute}

\begin{abstract}
For solving zero-sum games involving non-transitivity, a useful approach is to maintain a policy population to approximate the Nash Equilibrium (NE). Previous studies have shown that the Policy Space Response Oracles (PSRO) algorithm is an effective framework for solving such games. 
However, current methods initialize a new policy from scratch or inherit a single historical policy for Best Response (BR), missing the opportunity to leverage past policies to generate a better BR.
In this paper, we propose Fusion-PSRO, which employs Nash Policy Fusion to initialize a new policy for BR training. Nash Policy Fusion serves as an implicit guiding policy that starts exploration on the current Meta-NE, thus providing a closer approximation to BR. Moreover, it insightfully captures a weighted moving average of past policies, dynamically adjusting these weights based on the Meta-NE in each iteration. This cumulative process further enhances the policy population.
Empirical results on classic benchmarks show that Fusion-PSRO achieves lower exploitability, thereby mitigating the shortcomings of previous research on policy initialization in BR.
\end{abstract}
\end{frontmatter}

\section{Introduction} Zero-sum games, such as StarCraft~\cite{vinyals2019grandmaster,peng2017multiagent} and DOTA2~\cite{ye2020fullmoba}, involve strong non-transitivity, presenting unique challenges in game theory and artificial intelligence~\cite{Czarnecki2020RealWG}. Solving these non-transitive games often approximate a Nash Equilibrium (NE) by expanding a population of policies to handle diverse opponents. Traditional approaches like Fictitious Play~\cite{brown1951iterative} can converge to an NE by learning a set of policies, even in cyclic games like \textit{rock-paper-scissors}. For more complex games, the Policy Space Response Oracles (PSRO) algorithm~\cite{lanctot2017unified} has proven to be effective by iteratively learning the Best Responses (BRs) to the mixed strategies of opponents.

Previous research on PSRO has focused on improving policy diversity, training efficiency, and refining solution concepts. Diversity-enhancing variants such as Diverse PSRO~\cite{balduzzi2019diverse}, BD$\&$RD-PSRO~\cite{nieves2021modelling}, UDM-PSRO~\cite{liu2022unified}, and PSD-PSRO~\cite{yao2023policy} expand the set of policies to cover a wider policy space, thus improving robustness against diverse opponents. Pipeline-PSRO~\cite{mcaleer2020pipeline} accelerates training by parallelizing BRs, while approaches like $\alpha$-Rank~\cite{omidshafiei2019alpha} refine solution concepts to better approximate NE. However, few studies explore the impact of different policy initializations on PSRO performance. Current methods either initialize policies from scratch or inherit a single historical policy, despite PSRO generating numerous historical BRs that represent empirical responses to specific opponent strategies. This observation raises a key question: \textbf{Can the experience embedded in historical policies be leveraged to create stronger BRs?}

Many studies have shown that learning of policy ensembles can improve generalization capabilities \cite{sheikh2022dns,sharma2022deepevap}, suggesting that integrating historical BRs into PSRO could be beneficial. Therefore, our first idea is based on the fact that historical policies may contain useful experience for better BR training. To balance the contribution of each policy's experience, we leverage Meta-Strategy Solver as a trade-off mechanism. However, simply integrating these policies into PSRO does not guarantee optimal performance. Consequently, we propose fusing historical BRs into a single policy that achieves an ensemble-like effect and can be further fine-tuned to handle new opponents. This process of fusing historical policies aligns with the policy initialization of the BR in PSRO, forming the second key idea of our approach. 

In this paper, we propose the \textbf{Fusion-PSRO} framework, which enhances the traditional PSRO paradigm by focusing on improved policy initialization for better BR approximation. To achieve this, we present a novel method called \textbf{Nash Policy Fusion}, which serves as our specific implementation within the Fusion-PSRO framework.
This method uses all past BRs and then merges these into a single policy using weighted averaging according to the Meta-NE. Specifically, Nash Policy Fusion exhibits two key properties, that is, implicit guide policy and Nash weighted moving average. First, it provides a first-order approximation to a policy ensemble~\cite{munos2023Nash,rame2024rewarded}. 
Based on this, Nash Policy Fusion can be regarded as an implicit guide policy, which means that the policy jumps to the current Meta-NE and explores around it, thereby providing better data in the early stages of training~\cite{uchendu2023jump}. 
Second, from the perspective of the entire PSRO process, Nash Policy Fusion can be insightfully viewed as a special weighted moving average.
Each historical policy is optimized on the basis of the previous Meta-NE, and the new policy fuses the previously optimized BRs according to the current Meta-NE. This dynamic fusion of past BRs strengthens the entire policy population, promoting better convergence towards the NE.
This also aligns with the argument that weight averaging on top of adversarial training helps discover flatter minima and boosts robustness in both stochastic weight
averaging (SWA)~\cite{izmailov2018averaging} and exponential moving averages (EMA)~\cite{morales2024exponential,NEURIPS2021_fb4c4860}. 
Finally, our experiments validate that this plug-in initialization method of Fusion-PSRO can further reduce exploitability and achieve significantly better performance in diversity-enhanced methods.
\begin{figure}[htbp!]
    \includegraphics[width=\linewidth]{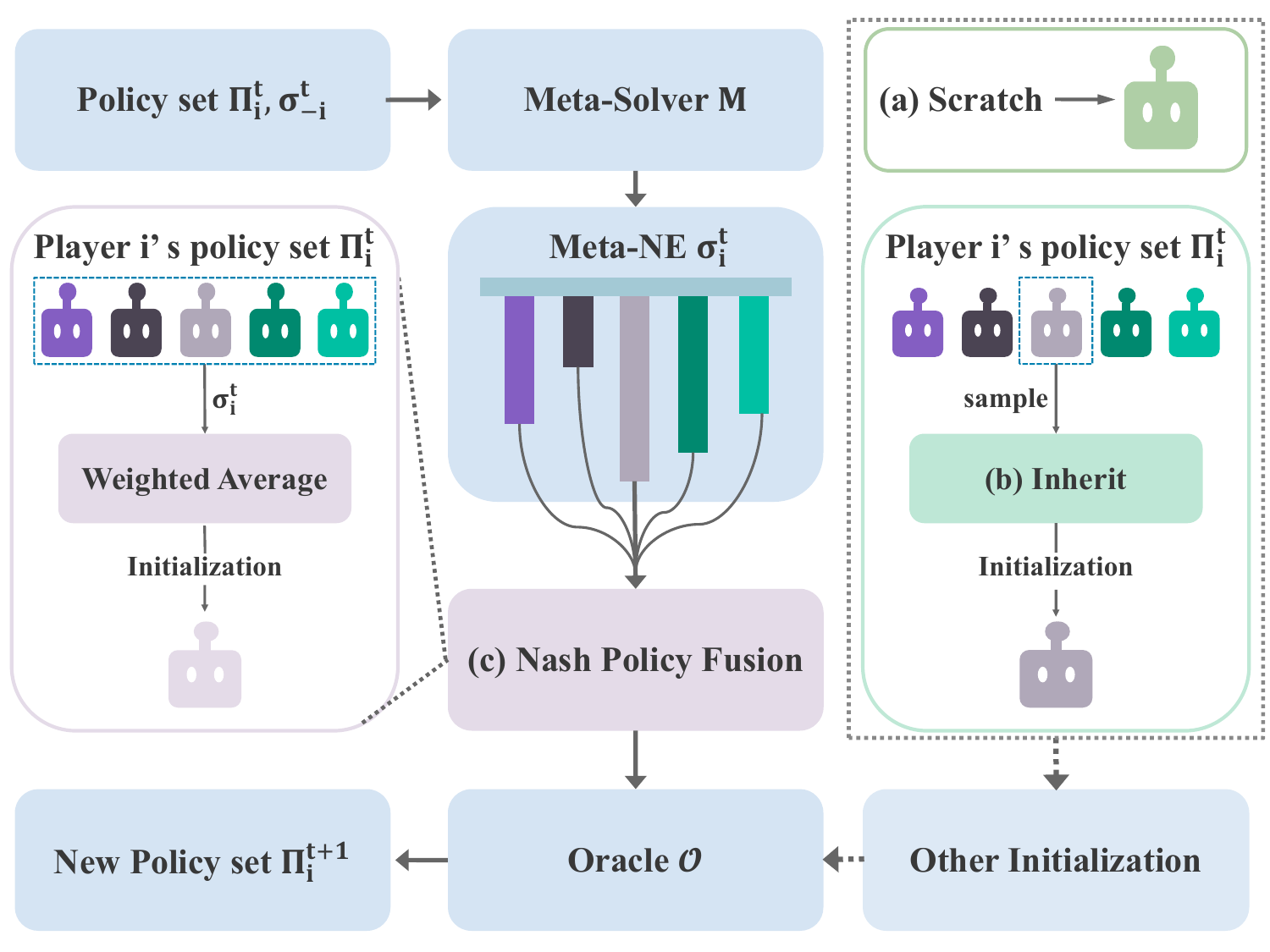}
    \caption{\textbf{Overview of the Fusion-PSRO Framework.} The diagram illustrates the various initialization methods for PSRO, especially Nash Policy Fusion.}
    \label{fusion-image}
    \addvspace{0.7\baselineskip}
\end{figure}
\section{NOTATIONS AND PRELIMINARY}
\textbf{Extensive-Form Game.} We analyze extensive-form games with perfect recall \cite{hansen2004dynamic}. In these games, players progress through a sequence of actions, each associated with a world state $w \in \mathcal{W}$. In an $N$-player game, the joint action space is $\mathcal{A} = \mathcal{A}_1 \times \cdots \times \mathcal{A}_N$. For player $i \in \mathcal{N} = \{1, \ldots, N\}$, the set of legal actions at world state $w$ is $\mathcal{A}_i(w) \subseteq \mathcal{A}_i$, and a joint action is $a = (a_1, \ldots, a_N) \in \mathcal{A}$. After players choose a joint action, the transition function $\mathcal{T}(w, a) \in \Delta^\mathcal{W}$ determines the probability distribution of the next world state $w'$. Upon transitioning from $w$ to $w'$ via joint action $a$, player $i$ observes $o_i = \mathcal{O}_i(w,a,w')$ and receives a reward $\mathcal{R}_i(w)$. The game ends after a finite number of actions when a terminal world state is reached.

\textbf{Metrics in Extensive-Form Game.} 
A history, denoted $h = (w^0, a^0, w^1, a^1, \ldots, w^t)$, is a sequence of actions and world states starting from the initial state $w^0$. 
An information set for player $i$, denoted $s_i$, is a sequence of that player's observations and actions up to that point: $s_i(h) = (a_i^0, o_i^1, a_i^1, \ldots, o_i^t)$. 
A player's strategy $\pi_i$ maps from an information set to a probability distribution over actions. A strategy profile $\pi$ is a tuple $(\pi_1, \ldots, \pi_N)$. Strategies of all players except $i$ are denoted $\pi_{-i}$. 
When a strategy $\pi_i$ is learned through reinforcement learning, it is referred to as a policy.
The expected value (EV) $v_i^{\pi}(h)$ for player $i$ is the expected sum of future rewards in history $h$ when all players follow strategy profile $\pi$. The EV for an information set $s_i$ is $v_i^{\pi}(s_i)$, and for the entire game, it is $v_i(\pi)$. In a two-player zero-sum game, $v_1(\pi) + v_2(\pi) = 0$ for all strategy profiles $\pi$. 
A Nash equilibrium (NE) is a strategy profile where no player can achieve a higher EV by deviating: $\pi^{*}$ is a NE if $v_i(\pi^{*}) = \max_{\pi_i}v_i(\pi_i, \pi^{*}_{-i})$ for each player $i$.
The \emph{exploitability} $e(\pi)$ of a strategy profile $\pi$ is the sum of improvements each player can make by deviating to their best response: $e(\pi) = \sum_{i \in \mathcal{N}} (v_i(\mathbb{BR}_i(\pi_{-i}), \pi_{-i}) - v_i(\pi_i, \pi_{-i}))$.
A BR strategy $\mathbb{BR}_i(\pi_{-i})$ for player $i$ maximizes exploitation of $\pi_{-i}$: $\mathbb{BR}_i(\pi_{-i}) = \arg\max_{\pi_i}v_i(\pi_i, \pi_{-i})$. An $\epsilon$-BR strategy $\mathbb{BR}^\epsilon_i(\pi_{-i})$ for player $i$ is at most $\epsilon$ worse than the BR: $v_i(\mathbb{BR}^\epsilon_i(\pi_{-i}), \pi_{-i}) \ge v_i(\mathbb{BR}_i(\pi_{-i}), \pi_{-i}) - \epsilon$. An $\epsilon$-Nash equilibrium ($\epsilon$-NE) is a strategy profile $\pi$ where $\pi_i$ is an $\epsilon$-BR to $\pi_{-i}$ for each player $i$.

\textbf{Normal-Form Game and Restricted Game.} 
For an iterative learning algorithm, we denote the set of all policies for player $i$ at iteration $t$ as $\Pi_i^t$.
A \emph{normal-form game} is a strategic representation of interactions where players make decisions simultaneously, which can be viewed as a degenerate case of extensive-form games with a single decision node for each player. An \emph{extensive-form} game induces a normal-form game where the legal actions for player $i$ are its deterministic strategies $X_{s_i \in \mathcal{I}_i} \mathcal{A}_i(s_i)$. These deterministic strategies are called pure strategies. A mixed strategy is a distribution over a player's pure strategies.
A \textit{restricted game}, as employed in PSRO, refers to a variant of a \emph{normal-form game} where players' strategies are limited to specific populations $\Pi^t_i$. 
A Meta-Strategy Solver (MSS) is a mechanism designed to optimize strategy selection and adaptation by leveraging historical performance and current game dynamics to enhance overall effectiveness, such as the Meta-Nash Solver.
A Meta-Nash Equilibrium (Meta-NE), denoted as $\sigma_i^t$, is the NE of the restricted game. It is a mixed strategy (a probability distribution) over the current policy population $\Pi_i^t$, which is computed by a MSS in each iteration.

\section{Related Work} 
\label{section-relatedwork}

\textbf{Ensembles and Fusion in Policy Learning.}
Policy ensembles and model fusion techniques provide complementary approaches to leverage historical knowledge in reinforcement learning. Ensembles of policies or value functions (e.g., Bootstrapped DQN~\cite{wu2020deep} and stochastic ensemble value expansion~\cite{buckman2018sample}) improve exploration and robustness by maintaining diverse strategies. Recent advances further optimize diversity through determinantal point processes~\cite{sheikh2022dns}.
However, directly deploying ensembles in PSRO incurs prohibitive computational costs, as adapting to evolving opponents requires repeated inference across all policies~\cite{song2023ensemble}.

Model fusion addresses this challenge by merging policies into a unified model. Weight averaging~\cite{munos2023Nash} and Fisher merging~\cite{matena2022merging} approximate ensemble benefits without runtime overhead, while mode connectivity~\cite{wortsman2022model} and parameter alignment~\cite{li2023deep} stabilize fused policies. In PSRO, methods like population parameter averaging~\cite{jolicoeur2023population} and the Rewarded Soup framework~\cite{rame2024rewarded} demonstrate that fused policies can retain diversity while reducing computational complexity. Our work adopts weight averaging as a lightweight fusion baseline, enabling efficient integration of historical policies into PSRO with minimal adaptation cost.

\textbf{Policy Initialization in PSRO.}
Policy initialization and iterative refinement are critical to the success of PSRO frameworks. Recent works have explored diverse strategies to initialize or update BR policies, yet face limitations in generality or efficiency. Mixed-Oracles~\cite{smith2021iterative} trains BR policies against the latest opponent strategy rather than the meta-NE distribution, rendering it highly sensitive to opponent distribution prediction errors.  Self-adaptive PSRO~\cite{li2024self} mitigates overfitting by blending the latest historical policy with a random policy. However, this approach underutilizes historical policies, as it excludes older strategies that could provide richer diversity. In contrast, our method fuses all historical policies via Nash-weighted averaging, including the initial random policy retained in the population, thereby maximizing experiential reuse.

Pre-trained models offer another initialization pathway but face domain-specific limitations. Grasper~\cite{li2024grasper} and pursuit-evasion solvers~\cite{li2023solving} leverage pre-trained policies tailored to specific games, yet these struggle in non-transitive settings where pre-trained policies may fail against diverse opponents. Fine-tuning BR policies from such models risks catastrophic deviation if the pre-trained models overfit to previous opponent strategies, undermining pre-trained knowledge. Our method dynamically assigns fusion weights based on historical policies' contributions and readjusts them when incorporating new strategies. This prevents overfitting to specific opponents while preserving opportunities to explore the optimal BR through anchored diversity.

\section{Fusion-PSRO}
\label{section-Fusionpsro}
We present \textbf{Fusion-PSRO}, an enhanced PSRO framework that improves policy initialization through \textbf{Nash Policy Fusion} – a novel method leveraging Nash-weighted averaging of historical BRs. Unlike existing PSRO variants that initialize policies from scratch or inherit, our approach unlocks latent synergies between historical policies via equilibrium-driven fusion, enabling more efficient exploration and stronger BR approximation.

\subsection{A Policy Fusion Framework for PSRO} 
Modern ensemble techniques in RL~\cite{sheikh2022dns,sharma2022deepevap} demonstrate that policy ensemble can enhance robustness and exploration by consolidating diverse strategies~\cite{liu2024neural}. However, directly applying these methods to PSRO proves prohibitive due to evolving opponent distributions and computational overhead. Our key insight lies in recognizing that the PSRO framework inherently provides an optimal fusion blueprint through its Meta-NE.

Building on insights from JSRL~\cite{uchendu2023jump}, we employ the following key assumptions and theoretical guarantees to support our approach:

\begin{remark}
\label{Quality of the Guide-Policy}
As proposed in JSRL, the guide-policy $\pi^g$ cover the states visited by the optimal policy $\pi^\star$:
\[
    \sup_{s, h} \frac{d_h^{\pi^\star}(\phi(s))}{d_h^{\pi^g}(\phi(s))} \leq C,
\]
where $s$ is parametrized by some feature mapping $\phi:\mathcal{S}\mapsto \mathbb{R}^d$, and $d_h^{\pi}$ represents the state distribution at step $h$ under policy $\pi$.
\end{remark}

\begin{remark}
\label{upper bound}
Under Remark~\ref{Quality of the Guide-Policy} and an appropriate choice of TrainPolicy and EvaluatePolicy, the $\mathsf{JSRL}$ algorithm guarantees a near-optimal bound up to a factor of $C \cdot \mathsf{poly}(H)$ for MDPs with general function approximation.
\end{remark}

Here, $H$ denotes the maximum steps in the game, and $\mathsf{poly}(H)$ is a polynomial function dependent on exploration methods such as $\epsilon$-greedy. Remark~\ref{Quality of the Guide-Policy} indicates that the guide policy $\pi^g$ aims to cover critical states in the feature space, even if suboptimal in action selection. Remarks \ref{Quality of the Guide-Policy} and \ref{upper bound} establish that a guide policy can reduce the sample complexity required for learning by covering key states of the optimal policy.

The fused policy $\pi_f$, derived from Nash-weighted fusion of historical policies, serves as an \textit{implicit guide policy}. By initializing training with $\pi_f$, we inherit the theoretical guarantees of JSRL, which reduces sample complexity through guided exploration. This contrasts with \textit{explicit guide policies} that require separate curriculum design. Our approach avoids such overhead. Subsequently, when policy fusion integrates diverse historical policies, the fused policy $\pi_{f}$ may contain a broader range of valuable states, improving exploration and potentially resulting in a lower coefficient $C$ compared to single-policy inheritance or scratch initialization. This ultimately contributes to a tighter near-optimal bound, as outlined in Remark~\ref{upper bound}.

\textbf{Fusion-PSRO Framework.} The Fusion-PSRO framework introduces policy fusion to enhance BR training. It comprises two key components: \emph{trade-off} and \emph{fusion}. The trade-off component creates a probability distribution based on diversity and utility across multiple policies, reflecting each policy's contribution to the population. This can be implemented using a MSS or dynamic learning coefficients during training. The fusion component merges these policies into a single policy, which is further fine-tuned. Techniques such as weighted averaging, policy distillation, or Fisher merging can be applied. 


As shown in Fig.~\ref{fusion-image}, Fusion-PSRO enhances PSRO by incorporating policy fusion into the initialization process. The Meta-NE $\sigma_i^t$ is first computed based on policy sets $\Pi_i^t$ and $\sigma_{-i}^t$. Three initialization methods are then proposed for player $i$: (a) \textbf{Scratch}: Initializes a new policy without leveraging historical policies; (b) \textbf{Inheritance}: Selects policies from $\Pi_i^t$ via $\sigma_i^t$ or reuses the latest BR; and (c) \textbf{Nash Policy Fusion}: Generates initial policies by $\sigma_i^t$-weighted aggregation of historical policies.  
Trained policies are then added to $\Pi_i^{t+1}$ after oracle optimization.

\begin{algorithm}[b!]
\caption{Initialization via Nash Policy Fusion}
\label{alg:fusion}
\begin{algorithmic}[1]
\State \textbf{Input:} Player $i$'s policy population $\Pi_i^t$ and parameters $\{\theta_\pi\}_{\pi \in \Pi_i^t}$, Meta-NE $\sigma_i^t$, fusion start iteration $c$.
\If{$t \ge c$}
    \State $\theta_{\pi_i^{t+1}} \gets \sum_{\pi \in \Pi_i^t} \sigma_i^t(\pi) \cdot \theta_\pi$.
\Else
    \State Sample $\pi \sim \sigma_i^t$ and initialize $\theta_{\pi_i^{t+1}} \gets \theta_\pi$.
\EndIf
\State \textbf{Output:} Initialized policy $\pi_i^{t+1}$.
\end{algorithmic}
\end{algorithm}

\begin{figure}[htbp!]
    \centering
    \includegraphics[width=0.5\textwidth]{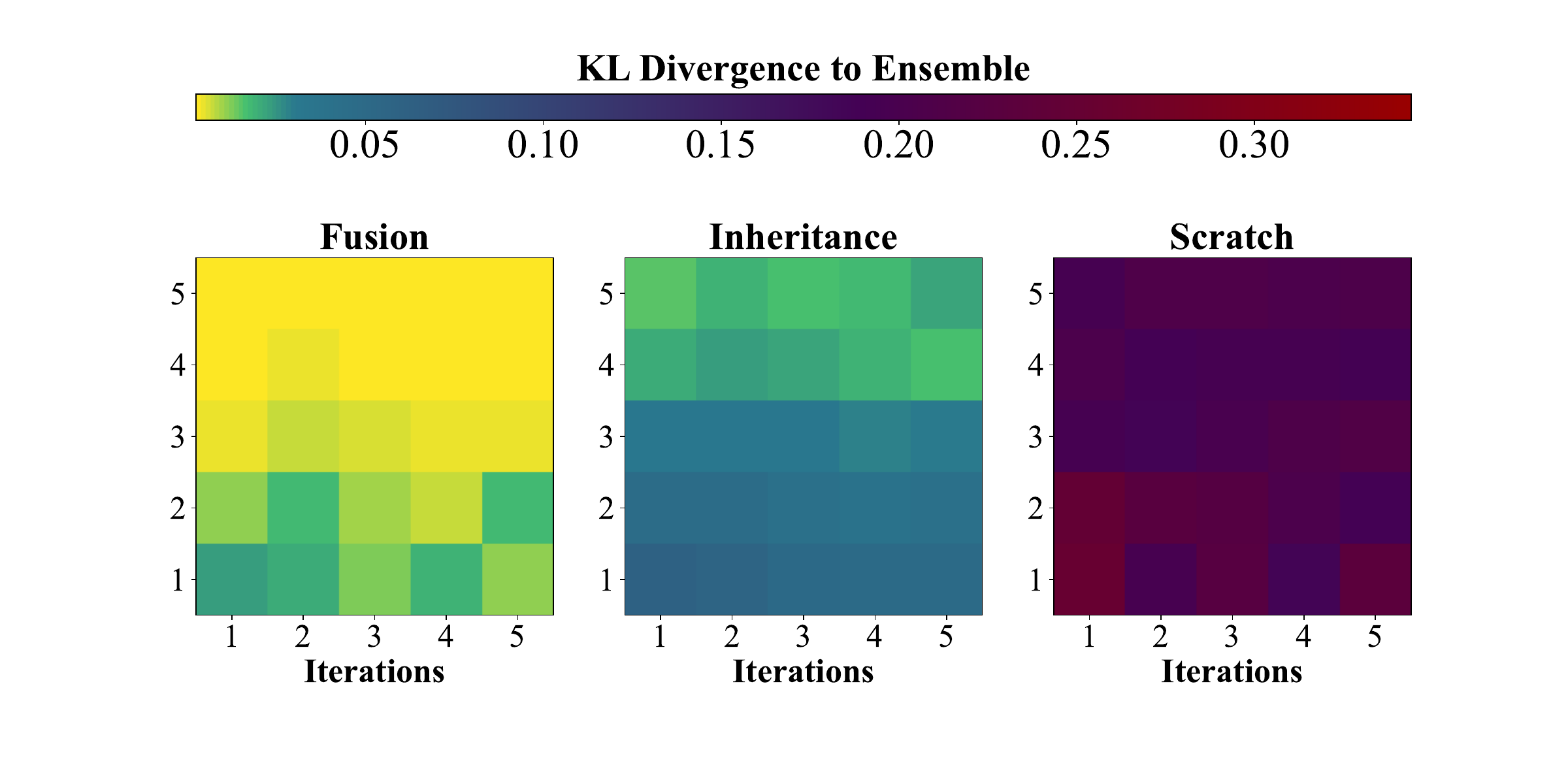}
    \vspace{-3\baselineskip}
    \caption{KL Divergence to ensemble for Fusion, Inheritance, and Scratch, with 25 tiles per method, each tile representing one training iteration.}
    \label{klcompare}
\end{figure}

\subsection{Nash Policy Fusion}
\label{nash policy fusion}
For the fusion method, while techniques like policy distillation and ensemble learning improve generalization and skill transfer, they come with high training costs. In contrast, weighted averaging (WA) offers a cost-effective alternative. WA directly averages the weights of base policies according to the Meta-NE probabilities, making it the preferred fusion approach in Fusion-PSRO, as detailed in Algorithm~\ref{Fusion-PSRO}. This method, termed \textbf{Nash Policy Fusion}, exhibits two key properties, i.e., implicit guide-policy and Nash weighted moving average.

\textbf{Implicit Guide-Policy.} WA provides a first-order approximation of a policy ensemble when the model weights are similar~\cite{wortsman2022model,munos2023Nash,rame2024rewarded}. Inspired by this, we measured the Kullback–Leibler (KL) divergence between the Nash policy ensemble and policies initialized via Nash Policy Fusion, inheritance, and scratch. Results in Fig.~\ref{klcompare} show that Nash Policy Fusion achieves the lowest KL divergence, closely approximating the Nash policy ensemble on Liar's Dice game. 

\begin{algorithm}[tbp!]
\caption{Fusion-PSRO}
\begin{algorithmic}[1]
\State \textbf{Input:} initial policy sets for all players $\Pi$
\State Compute utilities $U^{\Pi}$ for each joint $\pi \in \Pi$
\State Initialize Meta-NE $\sigma_i = \text{UNIFORM}(\Pi_i)$
\For{\emph{e} $\in \{1, 2, \dots\ ,N\}$}
    \For{\emph{player} $i \in \{1, 2, \dots, n\}$}
        \State Initialize $\pi^{t+1}_i$ via \textbf{Algorithm} \ref{alg:fusion} // Better Initialization
        \For{many episodes}
            \State Sample $\pi_{-i} \sim \sigma^t_{-i}$
            \State Train oracle $\pi^{t+1}_i$ over $\rho \sim (\pi^{t+1}_i, \pi_{-i})$
            \State \Comment{\textit{Train with RL against opponents sampled from $\sigma_{-i}$}}
        \EndFor        
        \State $\Pi^{t+1}_{i} = \Pi^{t}_{i} \cup \{\pi^{t+1}_i\}$
    \EndFor
    \State Compute payoffs for new policies against all others in $M^{t+1}$
    \State Compute a Meta-NE $\sigma$ from $M^{t+1}$
\EndFor
\State \textbf{Output:} current Meta-NE for each player
\end{algorithmic}
\label{Fusion-PSRO}
\end{algorithm}


This approximation enables Nash Policy Fusion to initialize new policies near the current Meta-NE, effectively "jump-starting" the value function with bootstrap data from high-value regions. The resulting exploration anchoring mechanism reduces exploration complexity by focusing early training on valuable states, as evidenced by: (a) Mirroring JSRL's benefits: Our method reduces exploration complexity; (b) Unlike explicit guide-policy or curriculum learning, the fused policy $\pi_f$ automatically prioritizes state visitation through Meta-NE, avoiding manual reward shaping; (c) BR Enhancement: Nash Policy Fusion bypasses random exploration phases and directly refines near-optimal strategies.

\begin{figure}[htbp!]
    \centering
    \begin{subfigure}[b]{0.23\textwidth}
        \centering
        \includegraphics[width=\textwidth]{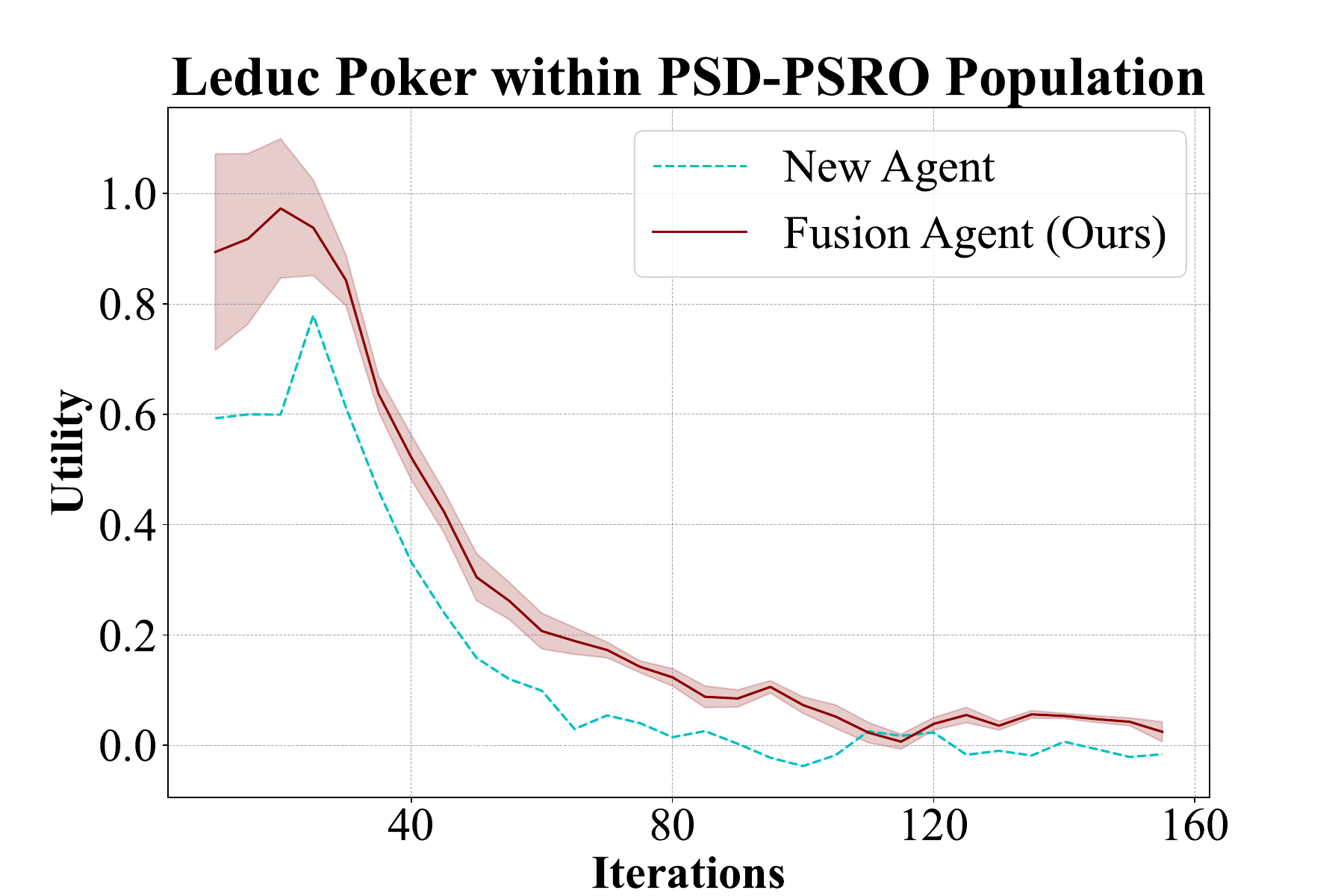}
        \caption{Utility from the meta-game between Fusion and Random Scratch in PSD-PSRO.}
        \label{fig:nash_reward_psd_psro}
    \end{subfigure}
    \begin{subfigure}[b]{0.23\textwidth}
        \centering
        \includegraphics[width=\textwidth]{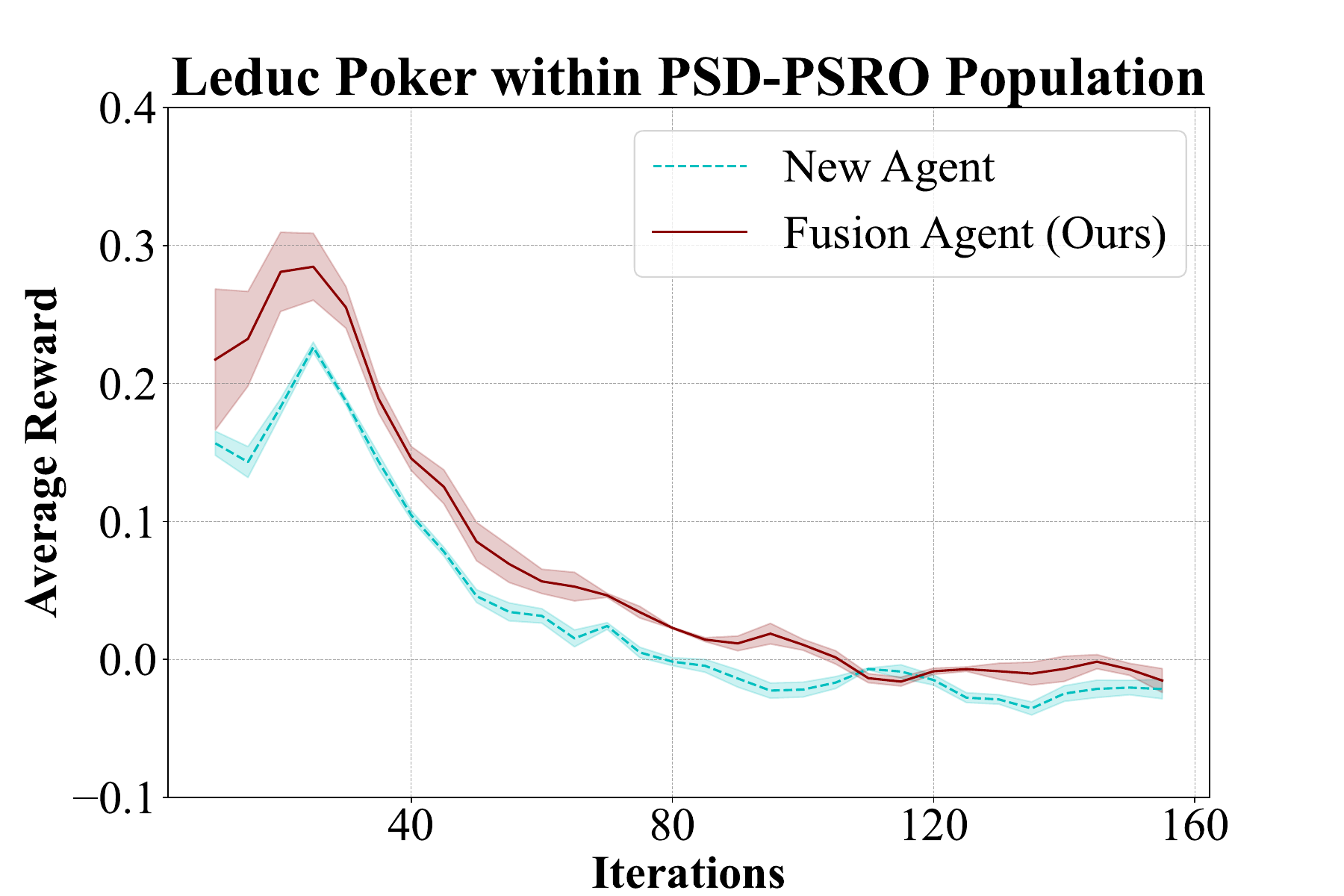}
        \caption{Reward from the full-game between Fusion and Random Scratch in PSD-PSRO.}
        \label{fig:reward_psd_psro}
    \end{subfigure}
    \caption{Under the condition of the same opponents, Average reward or utility between Fusion and Random Scratch across different iterations.}
    \label{compare_reward}
    \vspace{1\baselineskip}
\end{figure}

We evaluate the benefits of Nash Policy Fusion in achieving better approximation to BR for particular iterations. As shown in Fig.~\ref{compare_reward}, Nash Policy Fusion provides significant improvements in the PSD-PSRO setting, which enhances diversity. 
Nonetheless, across different initialization methods in Sec.~\ref{Ablations on Initialization Methods}, Nash Policy Fusion demonstrates superior performance, especially in early training stages, which facilitates exploration of high-reward states. In other words, Nash Policy Fusion explores the initialization policy within meta-NE. This implies that, when facing the current opponents, its reward is no worse than the rewards obtained by any of the current historical policies. As a result, it leads to a lower coefficient $C$ compared to inheritance and random initialization, thus achieving a more efficient training efficiency as show in fig.~\ref{liarsdice_reward}.

\textbf{Nash Weighted Moving Average.} While the implicit guide-policy in Nash Policy Fusion improves BR approximation, a key strength lies in its cumulative effect showed in Sec.~\ref{fusion_accumulate}. This is analogous to SWA, which maintains a uniform average of
checkpoints during the final stages of Stochastic Gradient Descent (SGD) or policy-from-scratch trajectories~\cite{nikishin2018improving}. Nash Policy Fusion, however, performs a weighted average of all previous BRs—insightfully viewed as iteration-level checkpoints. Unlike SWA's uniform weights, our method leverages the Meta-Nash
Solver to update the weights of BRs after each iteration, enabling more effective redistribution.

Such moving-average methods are beneficial because the loss landscape near a minimum is often asymmetric—sharp in some directions and flat in others. Stochastic Gradient Descent (SGD) tends to converge in sharp regions, while averaging iterates shifts the solution toward flatter areas~\cite{he2019asymmetric}. Weight averaging (WA), especially in adversarial or minimax settings, helps discover flatter minima and enhances robustness~\cite{morales2024exponential}.
In Nash Policy Fusion, each historical policy is optimized based on the Meta-NE at that time, and the new policy fuses the previously optimized BRs according to the current Meta-NE. The repeated process strengthens the entire policy population, thereby accelerating convergence toward the NE. Unlike EMA or uniform moving averages, which assign fixed weights to policies, Nash Policy Fusion dynamically adjusts each policy’s contribution based on the current Meta-NE, leading to better results in Sec.~\ref{fusion_accumulate}. 


\subsection{Nash Policy Fusion for PSROs}
After applying fusion-based initialization, we follow the standard PSRO training processes and objective functions. By incorporating the fusion approach, we introduce Nash Policy Fusion for PSRO, as detailed in Algorithm~\ref{Fusion-PSRO}. If the objective function uses regularization techniques from other PSRO variants or different training methods, the framework is referred to as Fusion-PSROs. This fusion mechanism can be integrated into nearly all PSRO variants, with experimental results presented in Section~\ref{section-experiment}.
Notably, Nash Policy Fusion operates as a plug-and-play module with remarkable flexibility across PSRO variants: 
\begin{itemize}[leftmargin=*,noitemsep]
    \item For \textbf{PSD-PSRO}~\cite{yao2023policy} and \textbf{PSRO$_{rN}$}~\cite{balduzzi2019open}, it suffices to execute Nash Policy Fusion upon the introduction of each new policy to the population.
    \item In \textbf{Pipeline-PSRO}~\cite{mcaleer2020pipeline}, the fusion process is triggered exclusively when incorporating the latest active policy into the meta-strategy.
\end{itemize}
This minimal adaptation requirement underscores the method's generalizability while preserving the computational simplicity of baseline algorithms.

\section{Experiments}
\label{section-experiment}
We evaluate the effectiveness of the Nash Policy Fusion in improving BR approximations and reducing exploitability. Our method is tested on various PSRO variants, including Pipeline-PSRO, PSRO$_{rN}$, BD$\&$RD-PSRO, Diverse-PSRO, and PSD-PSRO, collectively referred to as Fusion-PSROs. Baselines for comparison utilize the standard "inherited" initialization (i.e., from the latest policy), while our method uses Nash Policy Fusion. We benchmark the performance of these methods on both simple single-state games (e.g., Non-Transitive Mixture Game) and more complex extensive games (e.g., Leduc Poker, Goofspiel, and Liar's Dice). In nearly all benchmarks, Fusion-PSROs consistently exhibit lower \emph{exploitability}, as shown in Fig.~\ref{traj} and Fig.~\ref{fig:combined_exps}. 

\subsection{Experimental Results}
\begin{figure*}[tb!]
    \centering
    \begin{subfigure}[b]{0.9\textwidth}
        \centering
        \includegraphics[width=\linewidth]{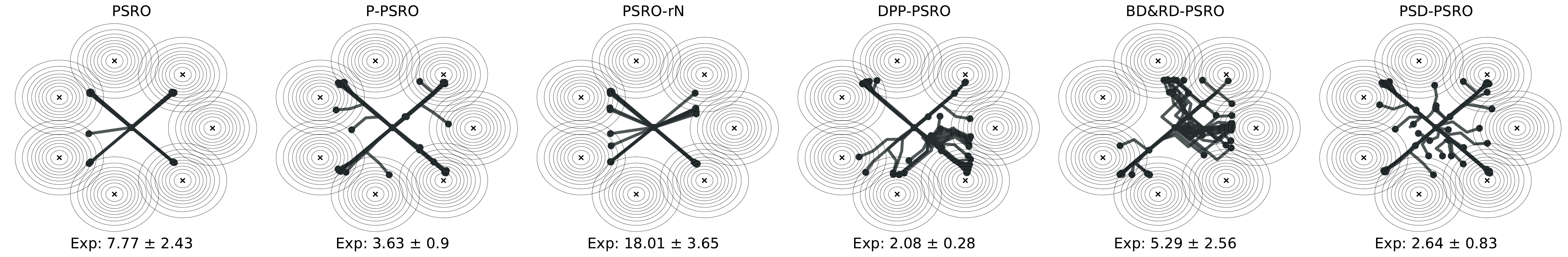}
        \caption{PSROs}
        \vspace{1.5\baselineskip}
        \label{trajectories_nofusion}
    \end{subfigure}
    \begin{subfigure}[b]{0.9\textwidth}
        \centering
        \includegraphics[width=\linewidth]{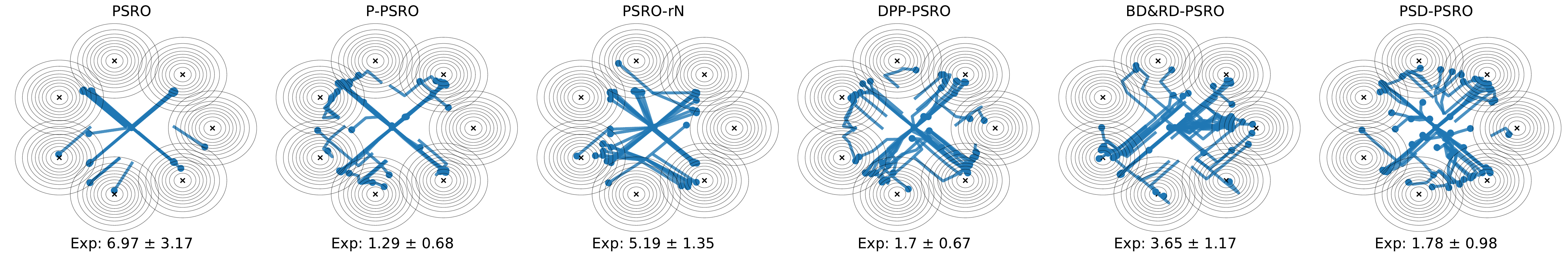}
        \caption{Fusion-PSROs}
        \label{trajectories_fusion}
    \end{subfigure}
    \caption{Training exploration trajectories on Non-Transitive Mixture Game. The final \emph{exploitability} ×100 (Exp) for each method is indicated at the bottom. And the more trajectories close to the centers of Gaussian, the higher the exploration efficiency of the algorithm.}
    \label{traj}
\end{figure*}

{\bf Non-Transitive Mixture Game.} 
This game features seven Gaussian humps distributed across a 2D plane, with each policy represented as a point on the plane. The payoff containing both non-transitive and transitive components is $\pi^T_i S \pi_{-i} + \frac{1}{2} \sum^7_{k=1} (\pi^k_i - \pi^k_{-i})$, where
\[
S = \begin{bmatrix}
0 & 1 & 1 & 1 & -1 & -1 & -1 \\
-1 & 0 & 1 & 1 & 1 & -1 & -1 \\
-1 & -1 & 0 & 1 & 1 & 1 & -1 \\
-1 & -1 & -1 & 0 & 1 & 1 & 1 \\
1 & -1 & -1 & -1 & 0 & 1 & 1 \\
1 & 1 & -1 & -1 & -1 & 0 & 1 \\
1 & 1 & 1 & -1 & -1 & -1 & 0 \\
\end{bmatrix}.
\]
Players must position themselves near the Gaussian centers to maximize performance.
As shown in Fig.~\ref{trajectories_fusion}, Fusion-PSROs generate denser exploration trajectories around these centers compared to standard PSROs (Fig.~\ref{trajectories_nofusion}), leading to better BR approximations. The initialization process in Fusion-PSROs results in trajectory starting points closer to the Gaussian centers, which helps traverse nearly all the centers during training, thus improving BR performance. Fusion-PSROs also exhibit lower \emph{exploitability} (Exp), indicating that policy fusion avoids unnecessary exploration. 

In contrast to traditional methods, where PSRO and PSRO$_{rN}$ are limited by the number of threads (1 compared to 4 in other PSRO variants), we increase the number of training episodes for these algorithms to maintain a balanced exploration effort.

\begin{figure*}[!tb]
    \centering
    \begin{subfigure}[b]{0.23\textwidth}
        \centering
        \includegraphics[width=\textwidth]{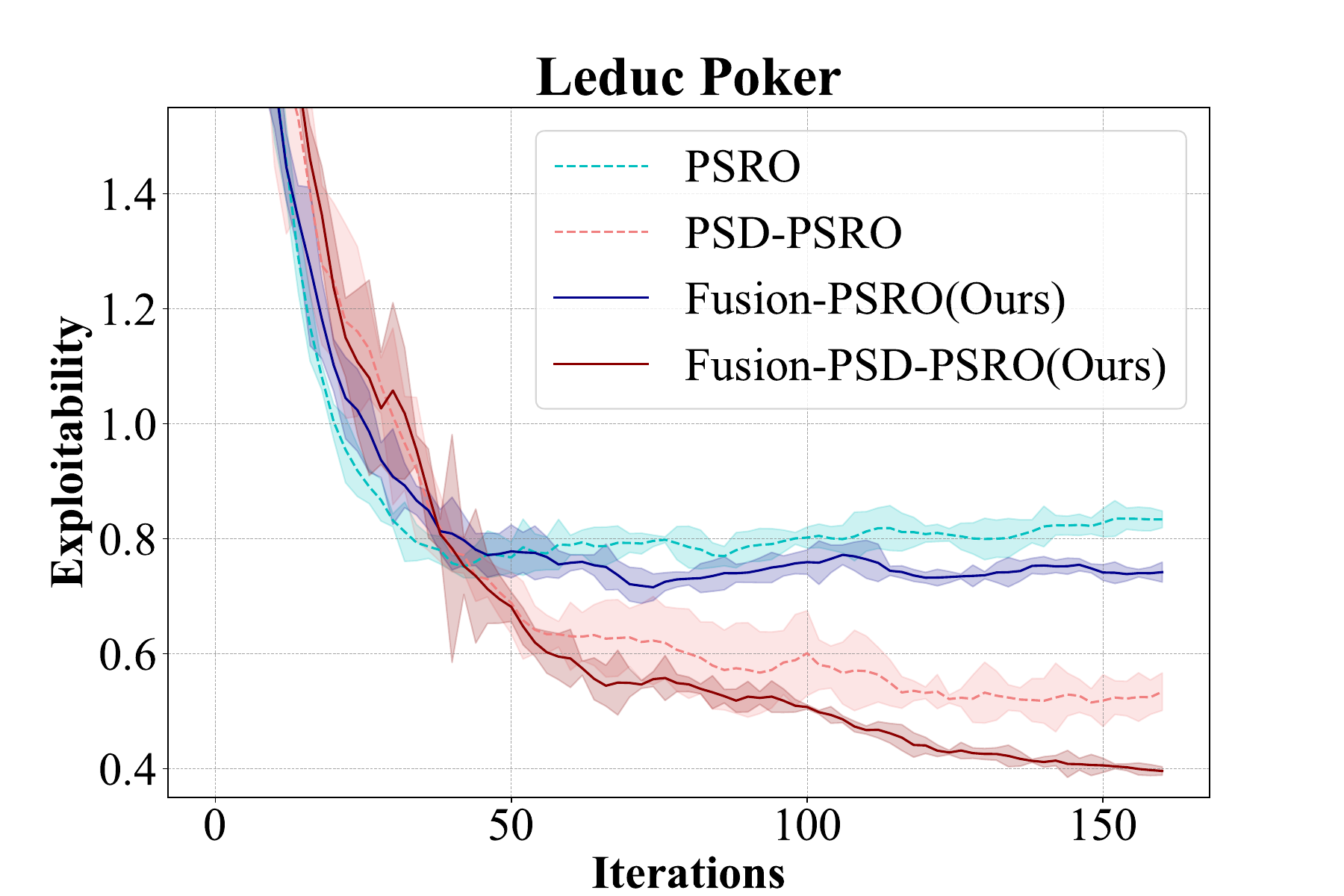}
        \caption{PSRO and PSD-PSRO: w/ Fusion vs w/o Fusion}
        \label{leduc_exp}
        \vspace{1\baselineskip}

    \end{subfigure}
    \hfill
    \begin{subfigure}[b]{0.23\textwidth}
        \centering
        \includegraphics[width=\textwidth]{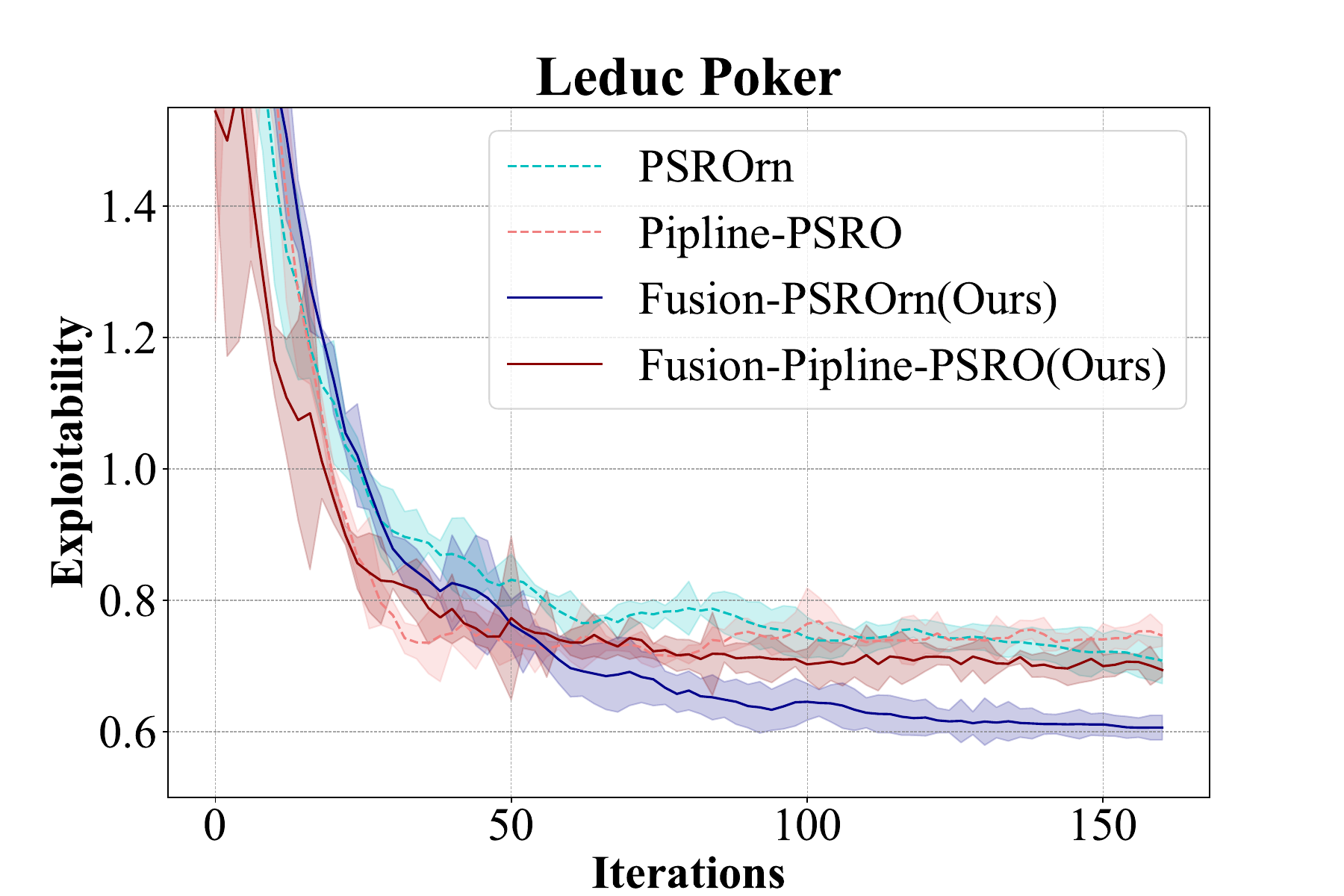}
        \caption{PSRO$_{rN}$ and P-PSRO: w/ Fusion vs w/o Fusion}
        \vspace{1\baselineskip}

        \label{leduc_exp_more_baselines}
    \end{subfigure}
    \hfill
    \begin{subfigure}[b]{0.23\textwidth}
        \centering
        \includegraphics[width=\textwidth]{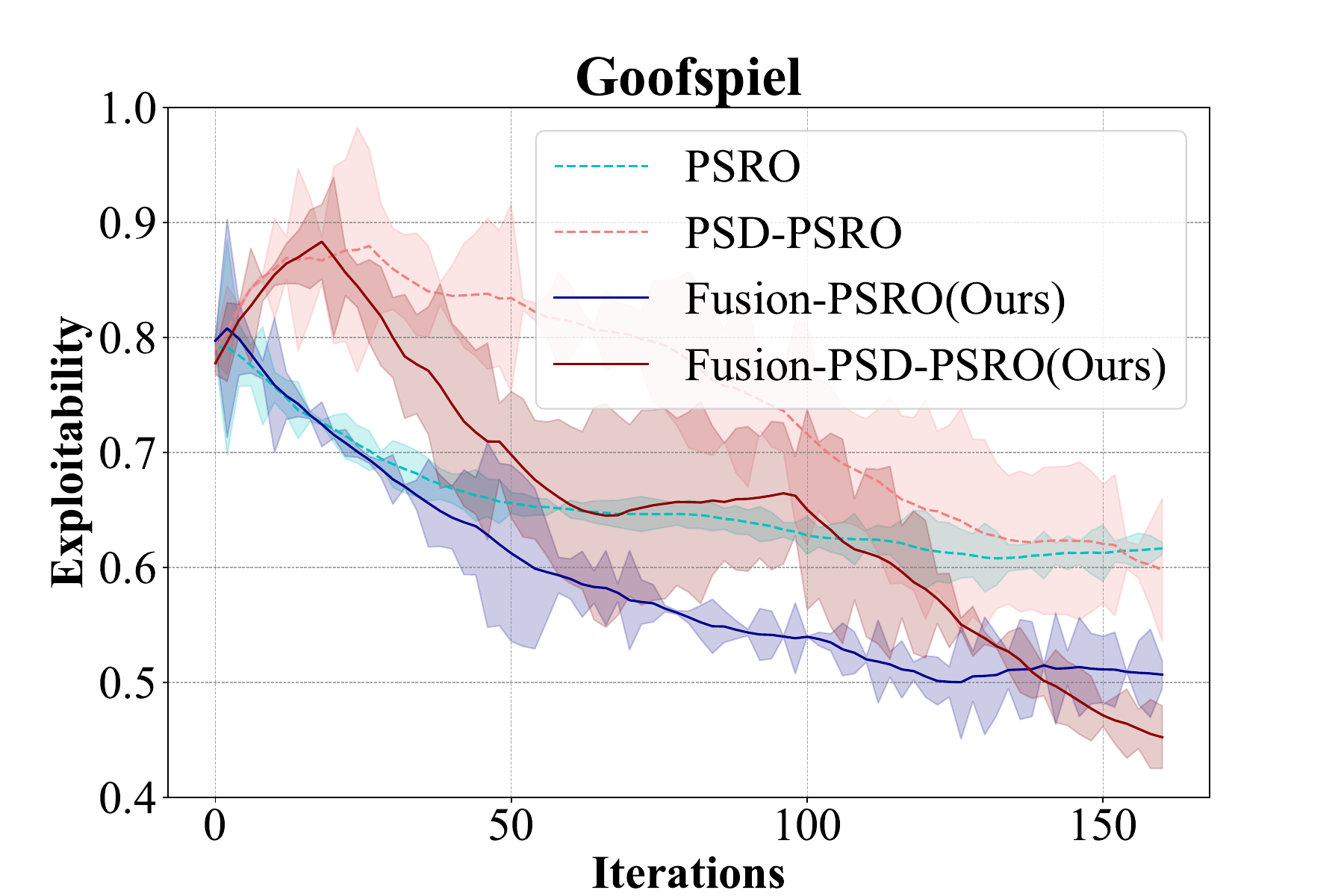}
        \caption{PSRO and PSD-PSRO: w/ Fusion vs w/o Fusion}
        \vspace{1\baselineskip}

        \label{Goofspiel_exp}
    \end{subfigure}
    \hfill
    \begin{subfigure}[b]{0.23\textwidth}
        \centering
        \includegraphics[width=\textwidth]{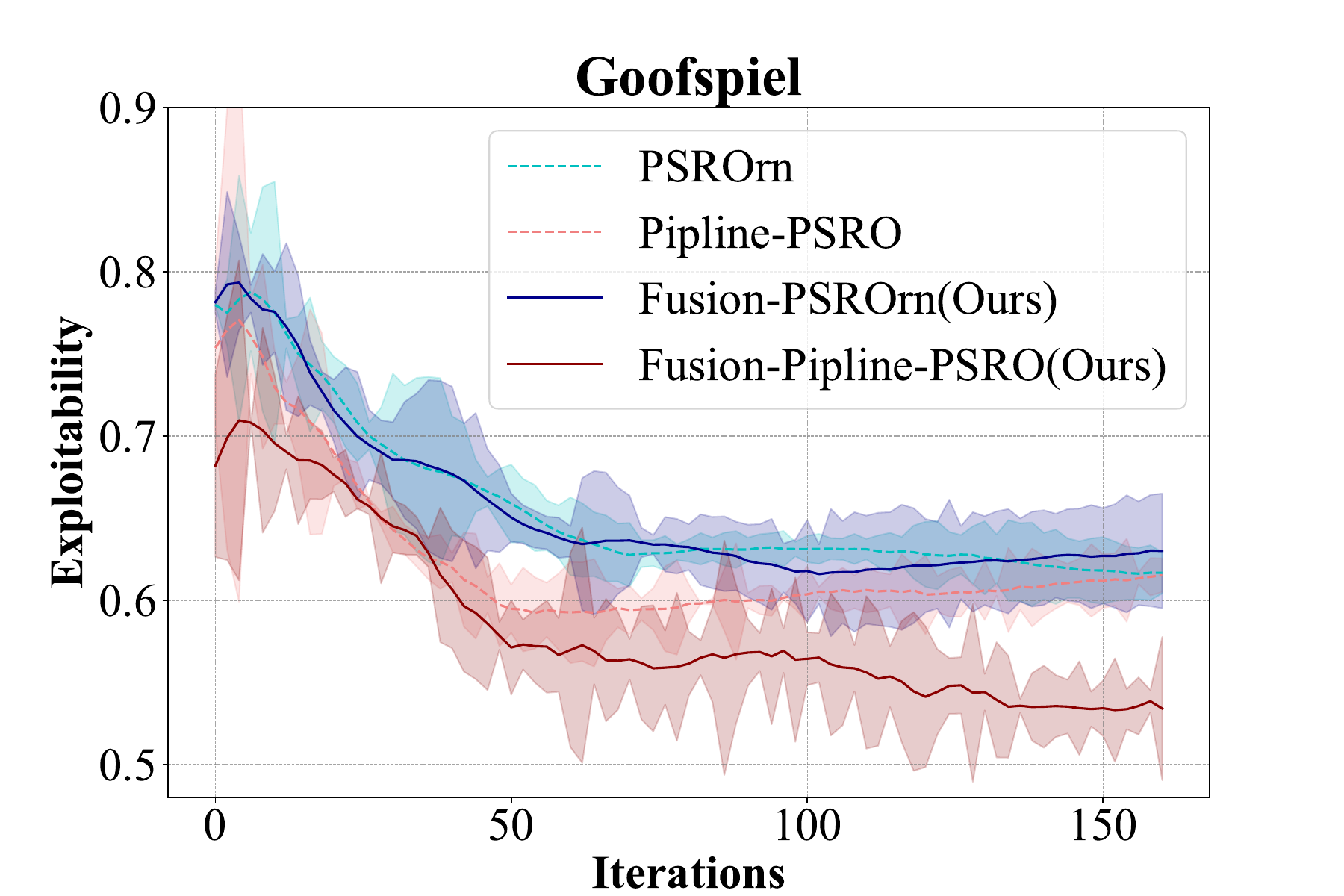}
        \caption{PSRO$_{rN}$ and P-PSRO: w/ Fusion vs w/o Fusion}
        \label{Goofspiel_exp_more_baselines}
        \vspace{1\baselineskip}

    \end{subfigure}
    \begin{subfigure}[b]{0.23\textwidth}
        \centering
        \includegraphics[width=\textwidth]{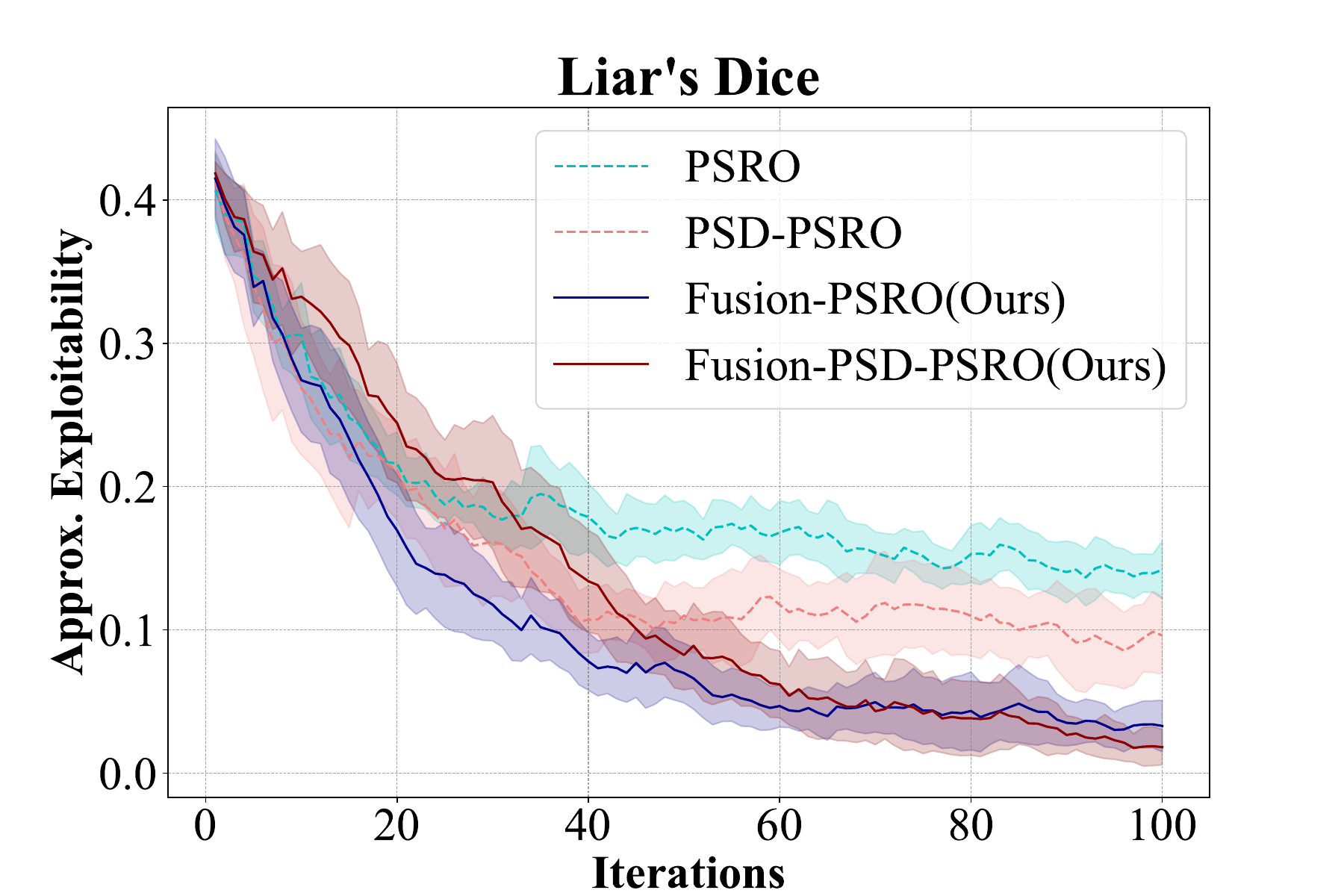}
        \caption{PSRO and PSD-PSRO: w/ Fusion vs w/o Fusion}
        \label{liarsdice_exp}
    \end{subfigure}
    \hfill
    \begin{subfigure}[b]{0.23\textwidth}
        \centering
        \includegraphics[width=\textwidth]{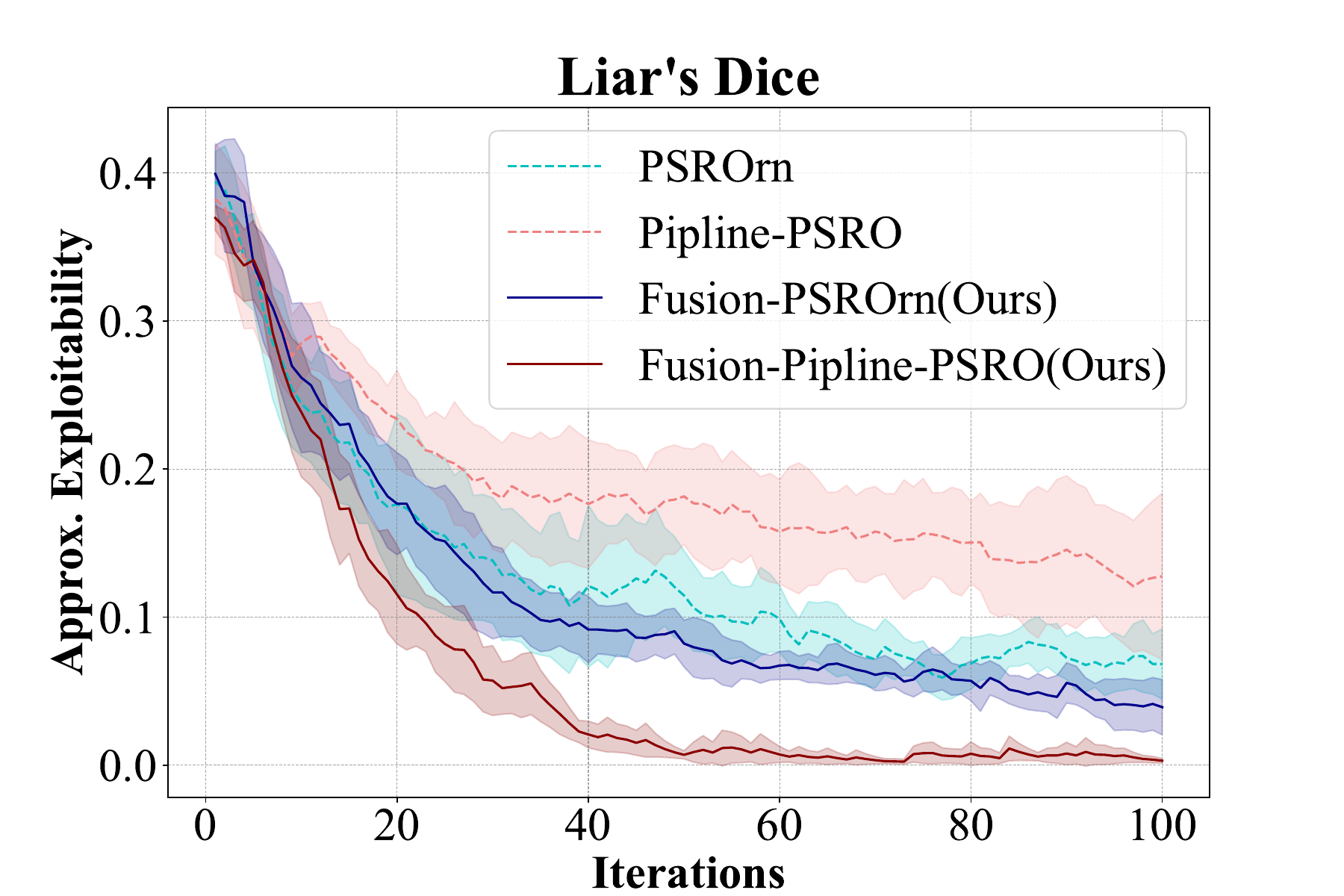}
        \caption{PSRO$_{rN}$ and P-PSRO: w/ Fusion vs w/o Fusion}
        \label{liarsdice_exp_more_baselines}
    \end{subfigure}
    \hfill
    \begin{subfigure}[b]{0.23\textwidth}
        \centering
        \includegraphics[width=\textwidth]{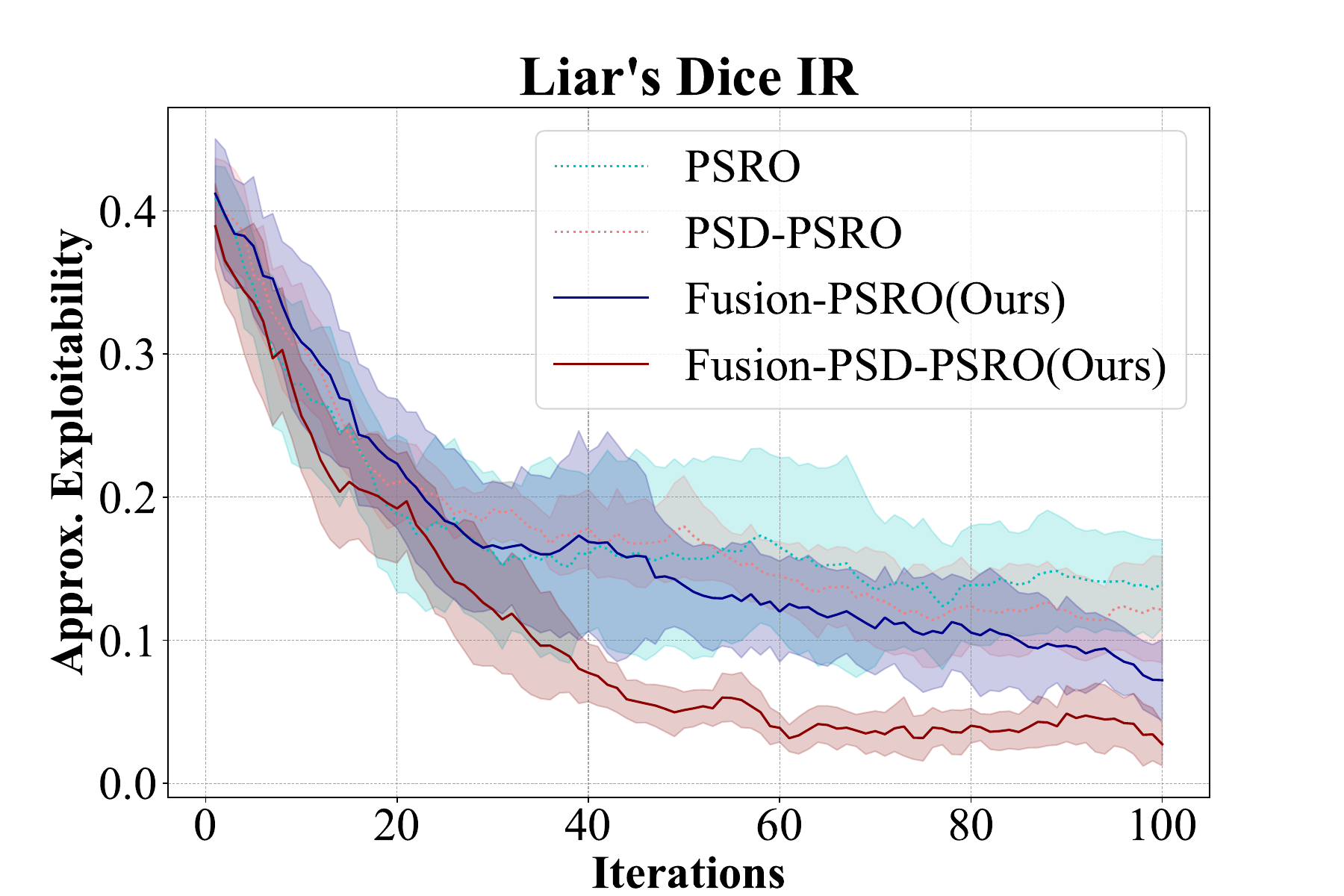}
        \caption{PSRO and PSD-PSRO: w/ Fusion vs w/o Fusion}
        \label{liarsdiceir_exp}
    \end{subfigure}
    \hfill
    \begin{subfigure}[b]{0.23\textwidth}
        \centering
        \includegraphics[width=\textwidth]{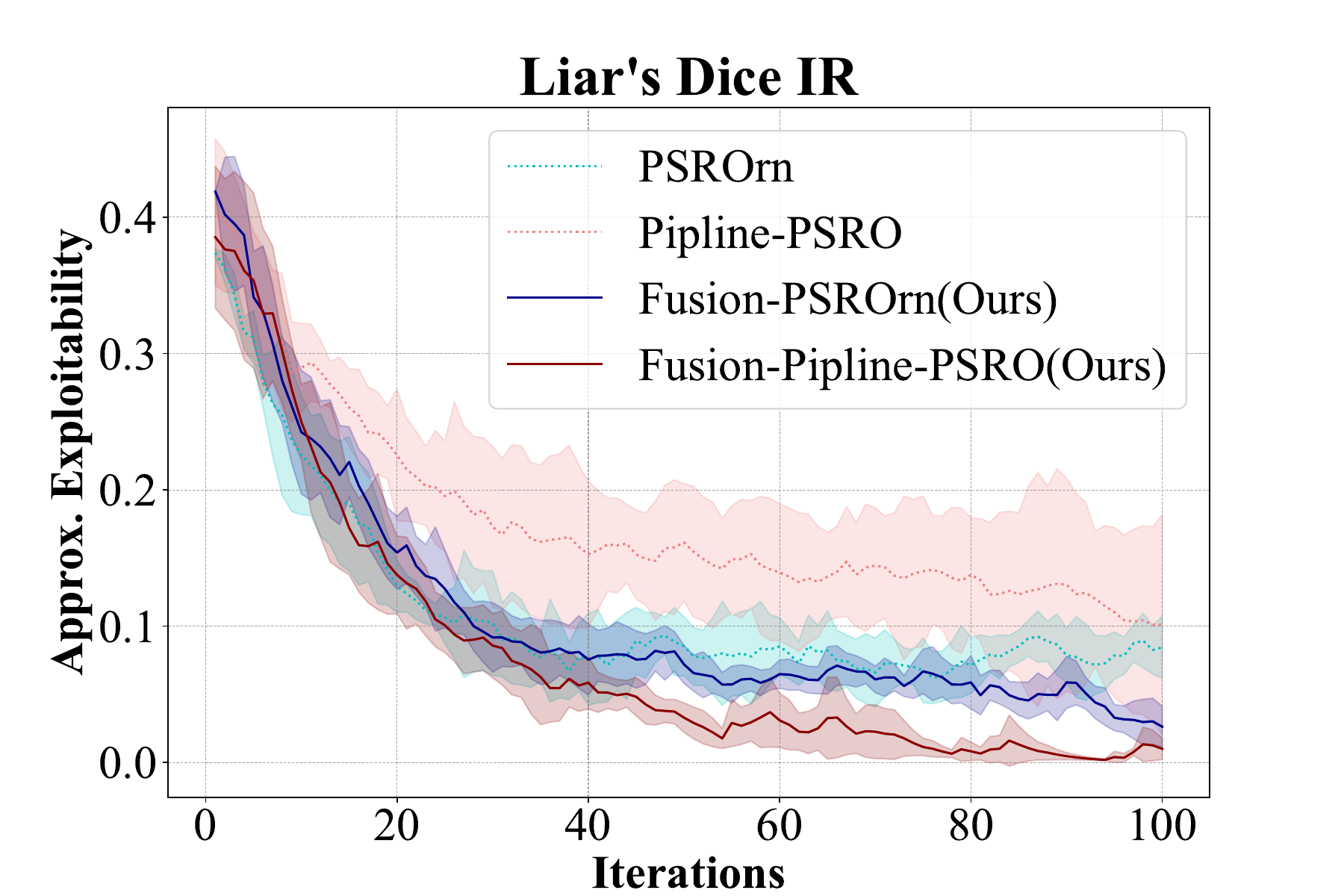}
        \caption{PSRO$_{rN}$ and P-PSRO: w/ Fusion vs w/o Fusion}
        \label{liarsdiceir_exp_more_baselines}
    \end{subfigure}

    \caption{
        \textbf{Impact of Fusion Module on Exploitability in Meta-NE.}
        Top row: Results on Leduc Poker and Goofspiel.
        Bottom row: Results on Liar's Dice and its imperfect information variant (Liar's Dice IR).
        Each subfigure compares the same algorithm \textbf{with fusion (w/ Fusion)} and \textbf{without fusion (w/o Fusion)}. In each subgraph, all hyperparameters are the same except for the initialization method. All experiments are run with at least three random seeds. The Nash Policy Fusion consistently reduces exploitability.
    }
    \label{fig:combined_exps}
\end{figure*}

{\bf Leduc Poker.} A simplified variant of poker \cite{southey2012bayes}, with a deck consisting of two suits, each containing three cards. Each player antes one chip, and a single private card is dealt to each player. 
Since Diverse-PSRO cannot scale to the RL setting and the code for BD\&RD-PSRO in complex games is unavailable, we compare other PSRO varients. As shown in Fig.~\ref{leduc_exp} and Fig.~\ref{leduc_exp_more_baselines}, both PSROs w/ Nash Policy Fusion are more effective at reducing \emph{exploitability} than their corresponding original version. Additionally, diversity-enhanced Fusion-PSD-PSRO outperforms Fusion-PSRO, highlighting the benefits of diversity in policy fusion. 

{\bf Goofspiel.} A large-scale, multi-stage simultaneous move game implemented in Openspiel~\cite{lanctot2017unified}.
Players compete through sequential bidding rounds using limited card inventories to win shifting prizes, creating complex strategy interactions. 
Fig.~\ref{Goofspiel_exp} and Fig.~\ref{Goofspiel_exp_more_baselines} show that Fusion-PSROs achieves lower \emph{exploitability} in strongly non-transitive games like Goofspiel, demonstrating the effectiveness of Nash Policy Fusion.

{\bf Liar's Dice.} A bluffing game where players make progressively higher bids about the count of a specific die face across all players~\cite{ferguson1991models}. The game alternates between making bids and challenging the veracity of the previous bid, leading to the loss of dice (or defeat) for incorrect challenges or bids.
As shown in Fig.~\ref{liarsdice_exp} and Fig~\ref{liarsdice_exp_more_baselines}, similar to Leduc Poker, by merging historical policies, Fusion-PSROs achieve lower \emph{approximate exploitability}~\cite{timbers2020approximate} than their corresponding original versions.

{\bf Liar's Dice IR.} A complex variant of Liar's Dice with imperfect recall. In Liar's Dice, players have perfect memory, allowing them to utilize all historical information when making decisions. However, in Imperfect Recall Liar's Dice, players' memory is limited to recalling only the most recent actions, which impacts their strategy choices. This limitation makes it even more critical to effectively leverage past experiences. As shown in Fig.~\ref{liarsdiceir_exp} and Fig.~\ref{liarsdiceir_exp_more_baselines}, our method enhances the performance of PSROs, achieving lower \emph{approximate exploitability}.

\subsection{Ablations on Initialization Methods}
\label{Ablations on Initialization Methods}
We compare four different policy initialization methods: normal initialization (random values from a normal distribution)~\cite{rumelhart1986learning}, orthogonal initialization (orthogonal weight matrices)~\cite{saxe2013exact}, Kaiming initialization (scaled weights)~\cite{he2015delving}, and inherited initialization (default in OpenSpiel, inheriting weights from previously trained policies)~\cite{lanctot2017unified}. Each initialization method is tested to generate approximate BRs (denoted as Normal-BR, Orthogonal-BR, Kaiming-BR, and Inherited-BR) during policy training within PSRO and PSD-PSRO. Simultaneously, we employ Nash Policy Fusion to produce Fusion-BR for comparison.
These BRs are not integrated into the population, so that its evolution continues to follow the baseline PSROs.

\begin{figure*}[!t]
    \centering
    \includegraphics[width=1\textwidth]{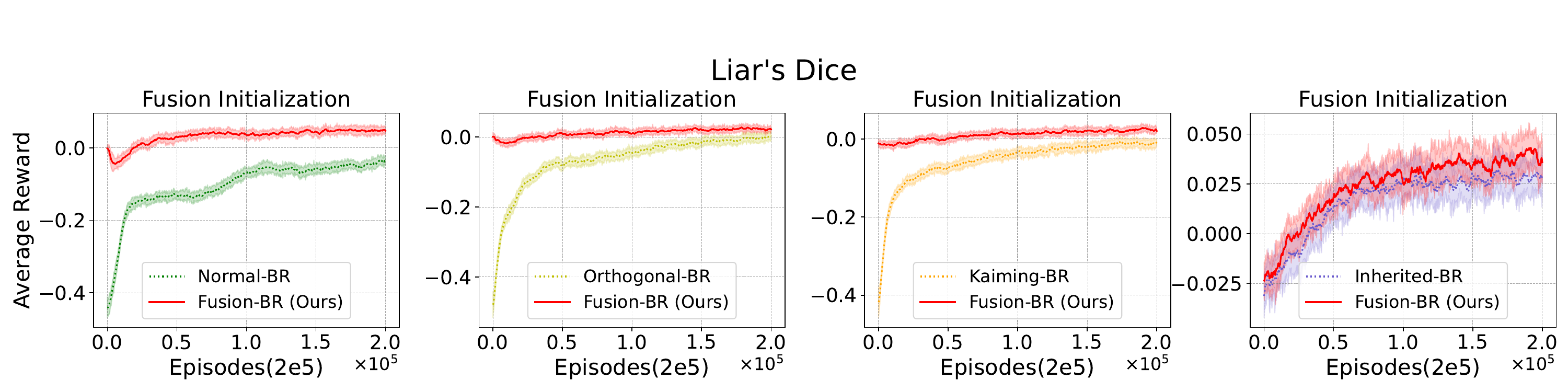}
    \caption{Average Reward averaged across iterations 71-80 for Approximate BRs. Fusion-BR (red) demonstrates consistently superior initial reward acquisition and faster convergence compared to conventional initialization baselines, while maintaining higher final performance. }
    \label{liarsdice_reward}
    \vspace{\baselineskip}
\end{figure*}

As shown in Fig.~\ref{liarsdice_reward}, Fusion-BR achieves higher initial rewards compared to the other initialization methods, which required more extensive training. In contrast, Fusion-BR converges to a higher average reward in fewer episodes. This suggests that Nash Policy Fusion not only accelerates convergence by initializing policies closer to optimal BRs but also reduces exploration redundancy. Consequently, Fusion-BR consistently outperforms almost all other initialization methods, demonstrating the effectiveness of using past policy knowledge to guide new policy. These results confirm that Fusion-PSROs, by incorporating Nash Policy Fusion, enable faster and better BR training.

\begin{figure}[htbp!]
    \centering
    \begin{subfigure}[b]{0.23\textwidth}
        \centering
        \includegraphics[width=\textwidth]{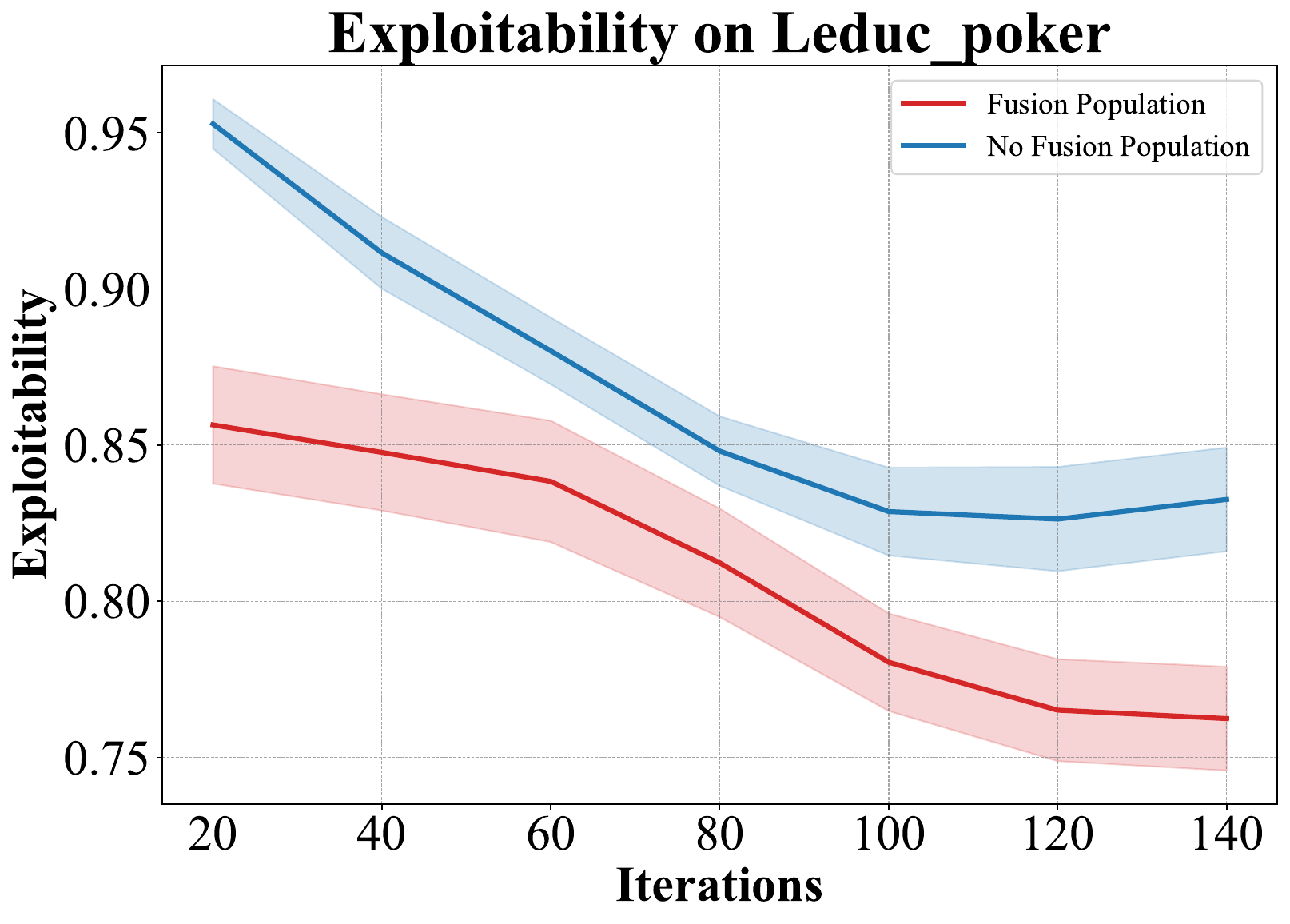}        \label{leduc_poker_exploitability_shaded}
    \end{subfigure}
    \hfill
    \begin{subfigure}[b]{0.23\textwidth}
        \centering   
        \includegraphics[width=\textwidth]{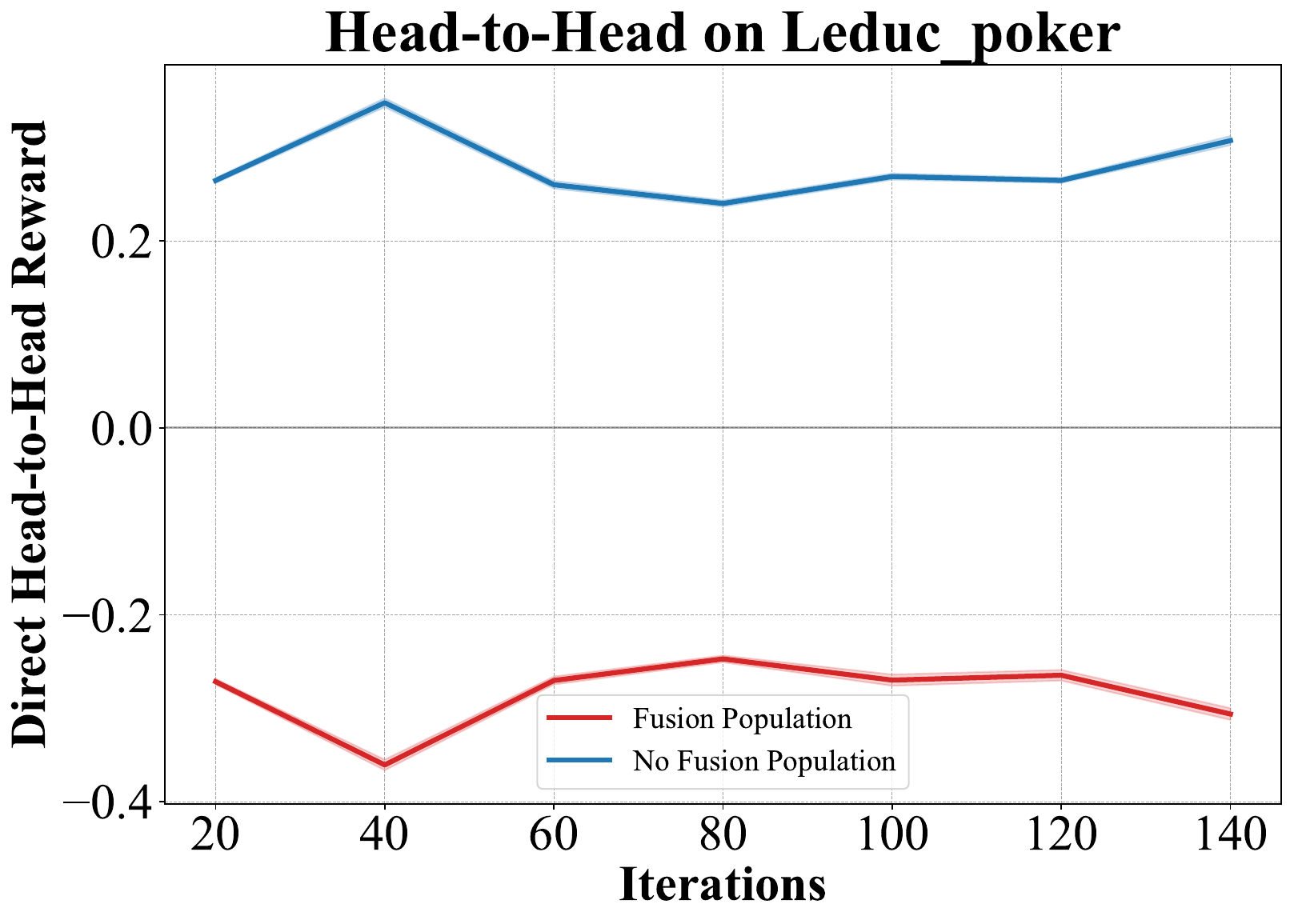}
        \label{head-to-head}
    \end{subfigure}
    \vspace{-1.5\baselineskip}
    \caption{Ablation Analysis of Nash Policy Fusion's Systemic Effects.}
    \vspace{1.4\baselineskip}
    \label{nw}
\end{figure}

\subsection{Ablations on Cumulative Effects}
\label{fusion_accumulate}
To isolate the cumulative effects of our method, we conduct a specialized ablation study. We compare a fusion-enabled population, trained with our framework, against a non-fusion population, trained with a standard baseline. As demonstrated in Figure~\ref{nw}, applying Nash Policy Fusion to the non-fusion population after training yields high head-to-head rewards (right), but this immediate advantage fails to reduce systemic exploitability (left). In contrast, the fusion-enabled population exhibits accelerated exploitability decay. This highlights that Nash Policy Fusion's true strength lies not in immediate gains, but in its ability to strategically guide the entire population’s evolution toward a more robust meta-Nash equilibrium.

\begin{figure}[htbp!]
    \centering
    \begin{subfigure}[b]{0.23\textwidth}
        \centering
        \includegraphics[width=\textwidth]{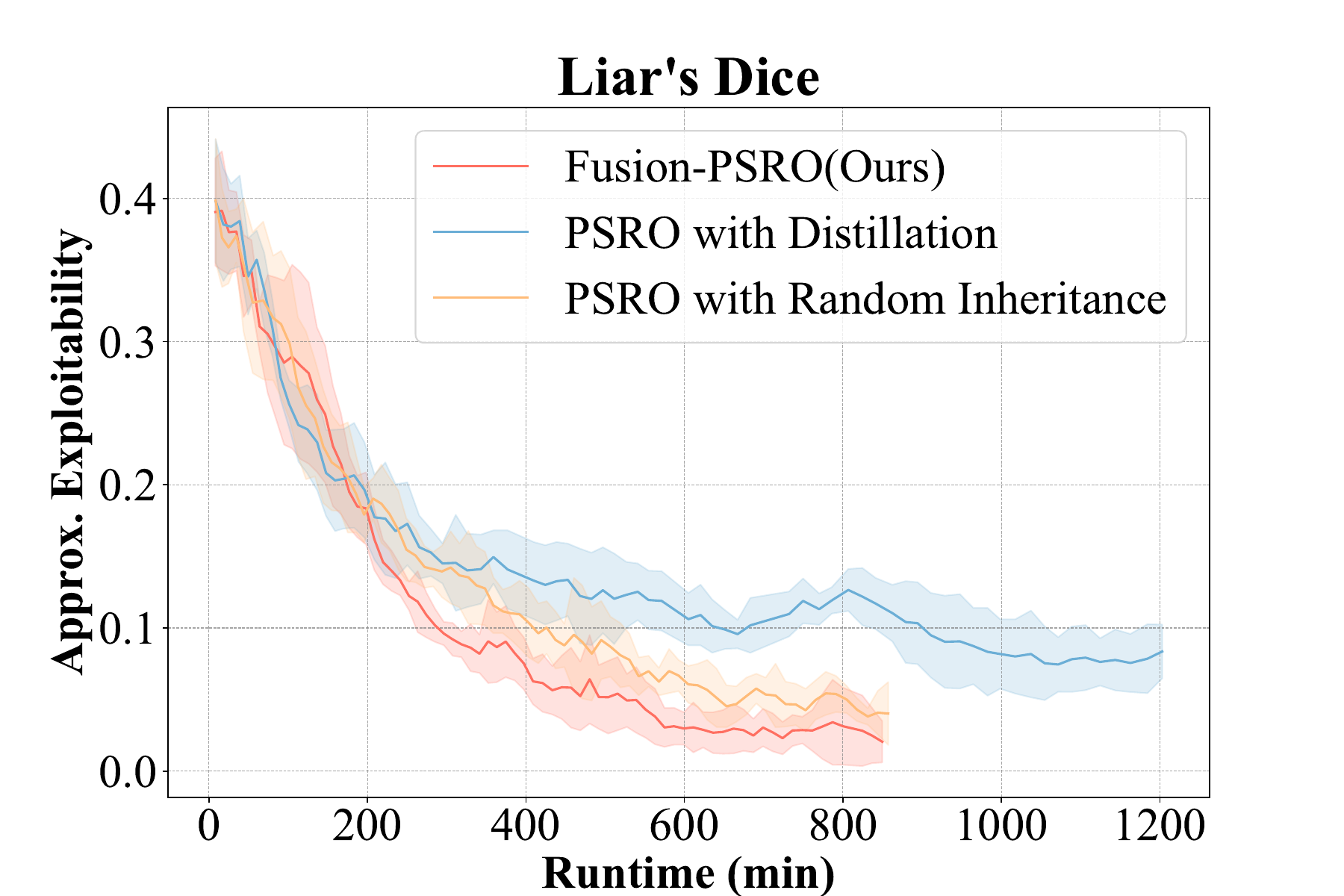}
        \caption{Exploitability of Nash Policy Fusion, Distillation, and Random Inheritance in PSRO.}
        \label{PSRO(randominheritance+distill+fusion)}
    \end{subfigure}
    \begin{subfigure}[b]{0.23\textwidth}
        \centering
        \includegraphics[width=\textwidth]{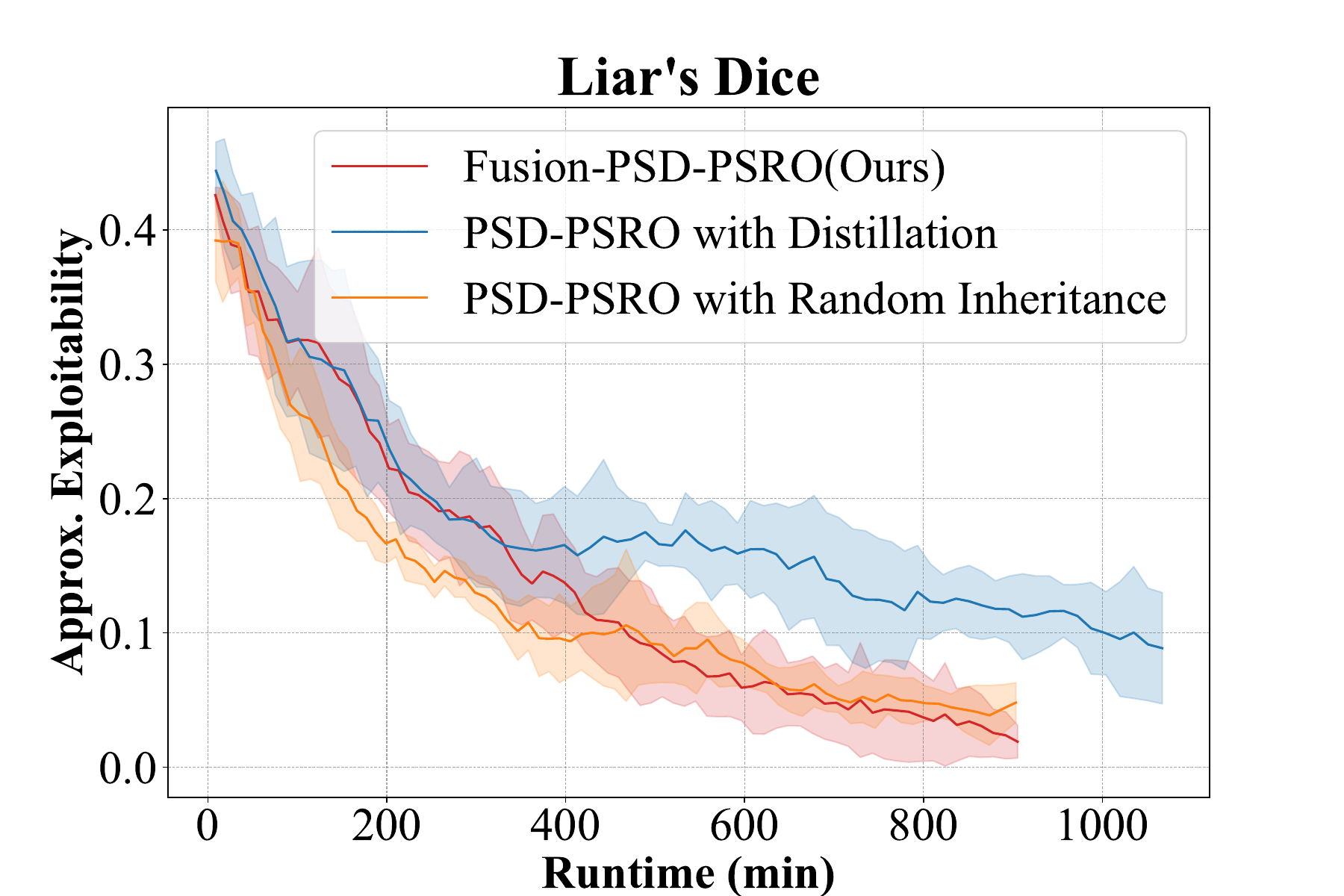}
        \caption{Exploitability of Nash Policy Fusion, Distillation, and Random Inheritance in PSD-PSRO.}
        \label{PSDPSRO(randominheritance+distill+fusion)}
    \end{subfigure}
    \caption{Ablation studies comparing Fusion, Distillation, and Random Inheritance in PSRO and PSD-PSRO over 1.5e7 episodes.}
    \label{ablation_studies_fig_1}
    \vspace{2\baselineskip}
\end{figure}
\begin{figure}[htbp!]
    \vspace{-10pt}
    \centering
    \begin{subfigure}[b]{0.23\textwidth}
        \centering        \includegraphics[width=\textwidth]{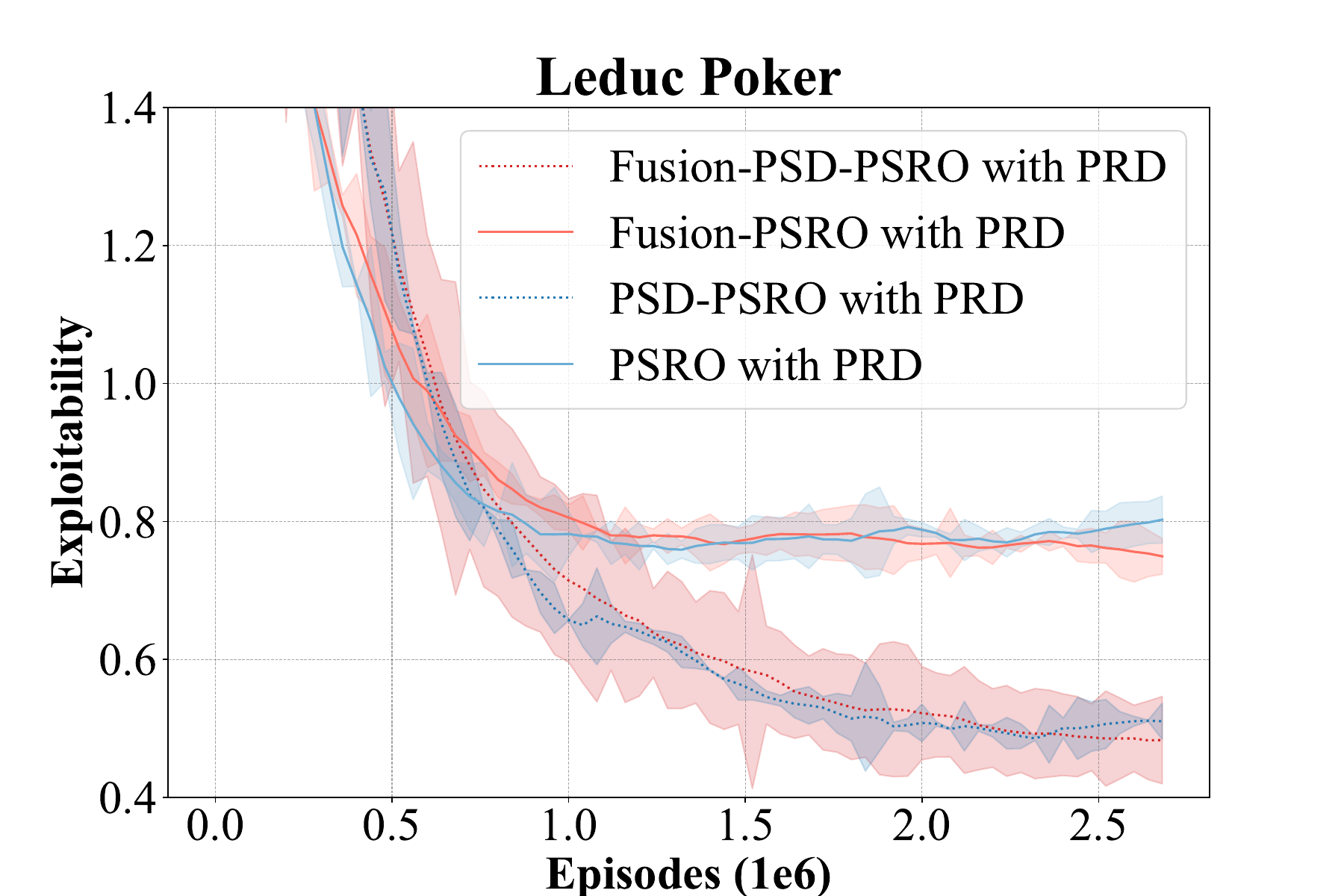}
        \caption{Comparisons between Nash Policy Fusion and its Version with PRD.}
        \label{r2_leduc_PRD}
    \end{subfigure}
    \begin{subfigure}[b]{0.23\textwidth}
        \centering   
        \includegraphics[width=\textwidth]{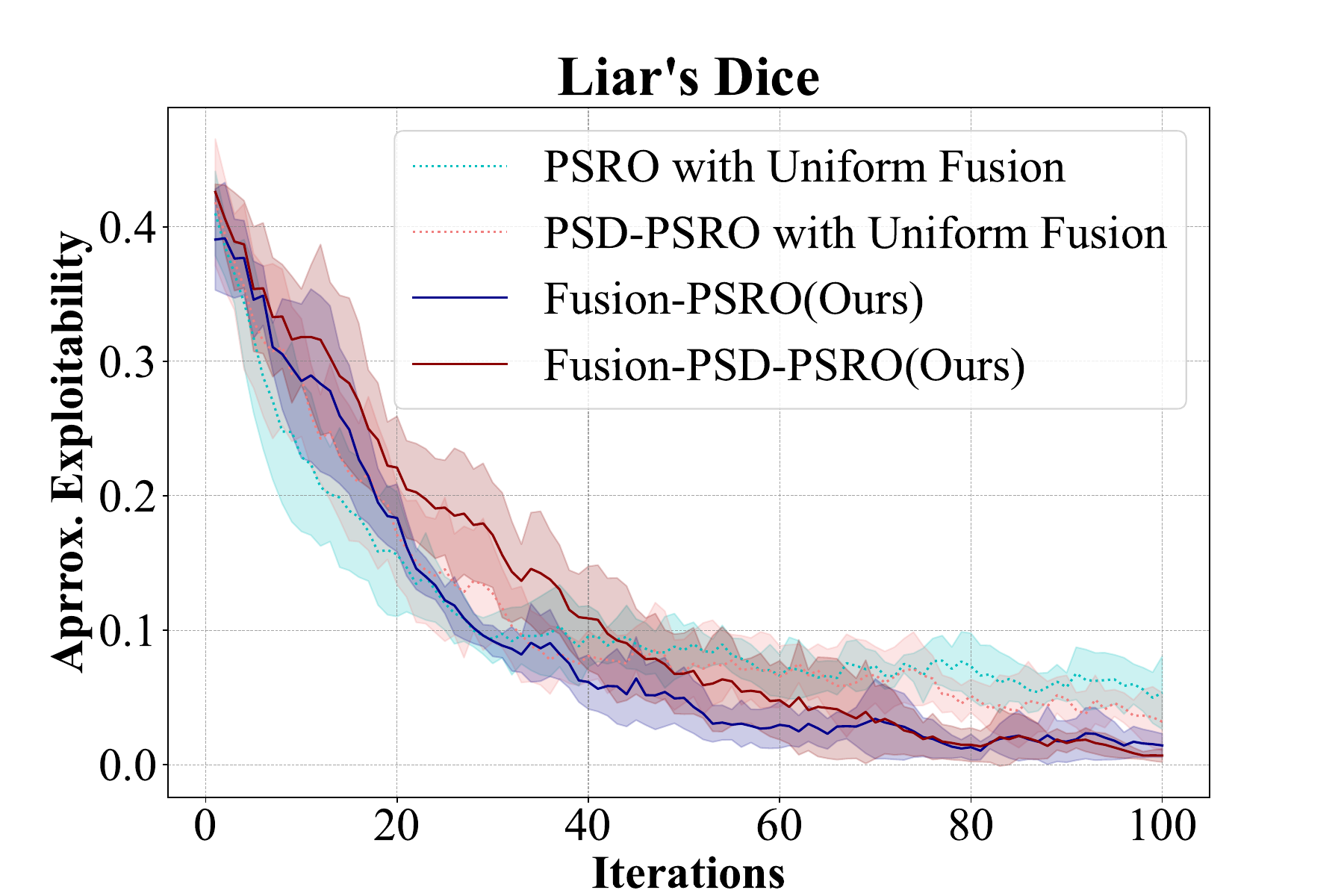}
        \caption{Comparisons between Fusion implementations with different fusion weights.}
        \label{r5_liarsdice_uniform_nash_iteration}
    \end{subfigure}
    \caption{Ablation studies comparing Fusion in PRD for Leduc Poker and the effect of different Fusion weights in Liar's Dice.}
    \vspace{1.5\baselineskip}
\end{figure}

\subsection{Ablations on MSS or Fusion Methods}
\label{ablations on liars_dice}
We conduct ablation studies to evaluate the effects of different MSS and policy fusion methods. Fig.~\ref{ablation_studies_fig_1} presents runtime comparisons on Liar's Dice between Nash Policy Fusion, policy distillation, and random inheritance. For policy distillation, we use the Nash policy ensemble approach, minimizing the policy distillation loss function $\mathbb{E}_{\pi}\left[\sum_{t} \mathrm{H}^{\times}\left(\pi\left(\tau_{t}\right) \| \pi_{\theta}\left(\tau_{t}\right)\right)\right]$~\cite{czarnecki2019distilling}. In the random inheritance method, we select a historical policy from the current Meta-NE based on its probability. By default, we inherit the latest policy from the previous iteration. Our results indicate that Nash Policy Fusion with weighted averaging achieves better performance on Liar’s Dice. In Fig.~\ref{r2_leduc_PRD}, we assess the impact of Projected Replicator Dynamics (PRD)~\cite{lanctot2017unified} as the MSS on Leduc Poker. We observe no performance improvement with policy fusion using PRD, possibly due to PRD focusing excessively on a small subset of policies. Finally, as demonstrated in Fig.~\ref{r5_liarsdice_uniform_nash_iteration}, while uniform weight moving averages show improvements over baseline methods, they fall short of the performance achieved by Nash Policy Fusion. This result means the advantage of Nash Weighted Moving Average, where dynamically adjusting weights based on the Meta-NE leads to lower \textit{exploitability}.

\subsection{Effect of the Fusion Starting Threshold ($c$)}
The fusion start threshold $c$ is empirically set based on NashConv convergence. Delaying fusion, e.g., until iteration $c=20$ in Goofspiel where NashConv stabilizes before decreasing, allows accumulation of diverse historical policies. Early fusion may use policies lacking necessary experience. Simpler environments like Leduc Poker, where NashConv drops quickly (e.g., after iteration 2), allow an earlier start ($c=2$).
Table~\ref{tab:fusion_start_concise} shows exploitability in Goofspiel at the 120th iteration for varied $c$. The $c=20$ achieves the lowest exploitability, demonstrating the benefit of an appropriately chosen delay.

\vspace{-0.8\baselineskip}
\subsection{Impact of the Number of Fused Policies}
We define a "useful" policy by its exploration of important states, assessed via meta-Nash probabilities. We experimented with fusing the top-$k$ most useful policies. Table~\ref{tab:top_k_policies_concise} shows exploitability on Leduc Poker (150th iteration) when varying $k$. Performance generally improves with $k$. Fusing all $n$ available policies yielded the best exploitability, significantly better than vanilla PSD-PSRO, indicating benefits from a broader range of historical experiences.

\begin{table}[htbp!]
\centering
\vspace{2\baselineskip}
\caption{Exploitability on Goofspiel (120th iteration) for different $c$.}
\begin{tabular}{lr}
\toprule
\textbf{Fusion Start Iteration ($c$)} & \textbf{Exploitability} \\
\midrule
0 (vanilla PSRO) & $0.607$ \\
2                & $0.628$ \\
10               & $0.577$ \\
20               & $\mathbf{0.527}$ \\
30               & $0.594$ \\
40               & $0.677$ \\
\bottomrule
\end{tabular}
\label{tab:fusion_start_concise}
\end{table}

\begin{table}[htbp!]
\centering
\vspace{0.5\baselineskip}
\caption{Exploitability on Leduc Poker (150th iteration) with varying numbers of top-ranked fused policies.}
\begin{tabular}{lr}
\toprule
\textbf{Number of Fused Policies (Top-$k$)} & \textbf{Exploitability} \\
\midrule
0 (vanilla PSD-PSRO) & $0.538$ \\
top1             & $0.469$ \\
top2             & $0.426$ \\
top3             & $0.435$ \\
top4             & $0.406$ \\
$n$ & $\mathbf{0.377}$ \\
\bottomrule
\end{tabular}
\label{tab:top_k_policies_concise}
\end{table}

\subsection{Comparison with Traditional Warm-Start Approaches}
Traditional warm-starting in RL, e.g., Warm-start Actor-Critic \cite{wang2023warm}, often uses ensembles to shape a new policy's value function. This can require many training batches and incur high inference costs (proportional to ensemble size), with difficulties in tuning the warm-start influence.

Nash Policy Fusion offers a more efficient warm start by initializing a new policy via Nash-weighted averaging of historical policy parameters. This avoids high computational costs (achieving $O(M)$ complexity for $M$ historical policies) and complex tuning. We set the warm-start proportion to $10\%$. Table~\ref{tab:goofspiel_comparison_concise} shows that on Goofspiel (5 cards), conventional warm-start PSRO ($0.639$) performs similarly to baseline PSRO ($0.626$). Fusion-PSRO, however, significantly reduces exploitability to $0.497$. Combined with PSD-PSRO, Fusion-PSD-PSRO achieves $0.427$. This highlights Nash Policy Fusion as a more effective and scalable warm-start solution.

\begin{table}[htbp!]
\centering
\vspace{\baselineskip}
\caption{Exploitability on Goofspiel (5 cards) comparing Fusion-PSRO with other algorithms and warm-start methods.}
\begin{tabular}{lr}
\toprule
\textbf{Algorithm} & \textbf{Exploitability} \\
\midrule
Warm-start (PSRO) & $0.639$ \\
PSRO              & $0.626$ \\
Fusion-PSRO       & $\mathbf{0.497}$ \\
PSD-PSRO          & $0.559$ \\
Fusion-PSD-PSRO   & $\mathbf{0.427}$ \\
\bottomrule
\end{tabular}
\label{tab:goofspiel_comparison_concise}
\end{table}




\section{Conclusions}
\label{Conclusions}
In this paper, we introduced the Fusion-PSRO framework, an enhancement to policy initialization within the PSRO paradigm using policy fusion techniques. Our approach, Nash Policy Fusion, leverages historical BRs by integrating them via Nash-weighted averaging, improving BR approximation without additional training overhead. Through extensive experiments across various games, we demonstrated that Nash Policy Fusion significantly enhances the performance of PSRO, achieving lower exploitability, particularly in diversity-enhanced PSROs. These results suggest that Nash Policy Fusion offers a simple and effective plug-in for improving both the initialization and overall performance of policy populations for PSRO.

\begin{ack}
This work was supported by guangdong Basic and Applied Basic Research Foundation 2024A1515030017 and 2024A1515011153, and also is supported by Wuhan Natural Science Foundation Exploratory Program (Chenguang Program) under Grant 2024040801020212.
\end{ack}
\bibliography{main}

@article{czarnecki2019distilling,
  title={Distilling Policy Distillation},
  author={Czarnecki, Wojciech Marian and Pascanu, Razvan and Osindero, Simon and Jayakumar, Siddhant M and Swirszcz, Grzegorz and Jaderberg, Max},
  journal={arXiv preprint arXiv:1902.02186},
  year={2019}
}

@article{balduzzi2019diverse,
  title={Diverse Population-Based Reinforcement Learning},
  author={Balduzzi, David and Racani{\`e}re, S{\'e}bastien and Martens, James and Foerster, Jakob and Tuyls, Karl and Graepel, Thore},
  journal={arXiv preprint arXiv:1901.08106},
  year={2019}
}

@inproceedings{ye2020fullmoba,
  title={Towards Playing Full MOBA Games with Deep Reinforcement Learning},
  author={Ye, Deheng and Chen, Guibin and Zhang, Wen and Chen, Sheng and Yuan, Bo and Liu, Bo and Chen, Jia and Liu, Zhao and Qiu, Fuhao and Yu, Hongsheng and others},
  booktitle={34th Conference on Neural Information Processing Systems (NeurIPS 2020)},
  year={2020},
  address={Vancouver, Canada}
}

@inproceedings{lanctot2017unified,
  title={A Unified Game-Theoretic Approach to Multiagent Reinforcement Learning},
  author={Lanctot, Marc and Zambaldi, Vinicius and Gruslys, Audrunas and Lazaridou, Angeliki and Tuyls, Karl and P{\'e}rolat, Julien and Silver, David and Graepel, Thore},
  booktitle={Advances in Neural Information Processing Systems},
  year={2017},
  address={Long Beach, CA, USA}
}

@inproceedings{mcaleer2020pipeline,
  title={Pipeline PSRO: A Scalable Approach for Finding Approximate Nash Equilibria in Large Games},
  author={McAleer, Stephen and Lanier, John and Fox, Roy and Baldi, Pierre},
  booktitle={Advances in Neural Information Processing Systems},
  year={2020},
  address={Vancouver, Canada},
  email={smcaleer@uci.edu, jblanier@uci.edu, royf@uci.edu, pfbaldi@ics.uci.edu},
  url={https://github.com/JBLanier/pipeline-psro}
}

@inproceedings{yao2023policy,
  title={Policy Space Diversity for Non-Transitive Games},
  author={Yao, Jian and Liu, Weiming and Fu, Haobo and Yang, Yaodong and McAleer, Stephen and Fu, Qiang and Yang, Wei},
  booktitle={Advances in Neural Information Processing Systems},
  year={2023},
  email={haobofu@tencent.com}
}

@inproceedings{nieves2021modelling,
  title={Modelling Behavioural Diversity for Learning in Open-Ended Games},
  author={Nieves, Nicolas Perez and Yang, Yaodong and Slumbers, Oliver and Mguni, David Henry and Wen, Ying and Wang, Jun},
  booktitle={Advances in Neural Information Processing Systems},
  year={2021},
  email={yaodong.yang@outlook.com}
}

@article{matena2022merging,
  title={Merging models with fisher-weighted averaging},
  author={Matena, Michael S and Raffel, Colin A},
  journal={Advances in Neural Information Processing Systems},
  volume={35},
  pages={17703--17716},
  year={2022}
}

@article{jolicoeur2023population,
  title={Population parameter averaging (papa)},
  author={Jolicoeur-Martineau, Alexia and Gervais, Emy and Fatras, Kilian and Zhang, Yan and Lacoste-Julien, Simon},
  journal={arXiv preprint arXiv:2304.03094},
  year={2023}
}

@article{li2023deep,
  title={Deep Model Fusion: A Survey},
  author={Li, Weishi and Peng, Yong and Zhang, Miao and Ding, Liang and Hu, Han and Shen, Li},
  journal={ArXiv preprint ArXiv:2309.15698},
  year={2023}
}

@inproceedings{wortsman2022model,
  title={Model soups: averaging weights of multiple fine-tuned models improves accuracy without increasing inference time},
  author={Wortsman, Mitchell and Ilharco, Gabriel and Gadre, Samir Ya and Roelofs, Rebecca and Gontijo-Lopes, Raphael and Morcos, Ari S and Namkoong, Hongseok and Farhadi, Ali and Carmon, Yair and Kornblith, Simon and others},
  booktitle={International conference on machine learning},
  pages={23965--23998},
  year={2022},
  organization={PMLR}
}

@inproceedings{he2015delving,
  title={Delving Deep into Rectifiers: Surpassing Human-Level Performance on ImageNet Classification},
  author={He, Kaiming and Zhang, Xiangyu and Ren, Shaoqing and Sun, Jian},
  booktitle={Proceedings of the IEEE international conference on computer vision},
  pages={1026--1034},
  year={2015}
}

@article{saxe2013exact,
  title={Exact solutions to the nonlinear dynamics of learning in deep linear neural networks},
  author={Saxe, Andrew M and McClelland, James L and Ganguli, Surya},
  journal={arXiv preprint arXiv:1312.6120},
  year={2013}
}

@article{rumelhart1986learning,
  title={Learning representations by back-propagating errors},
  author={Rumelhart, David E and Hinton, Geoffrey E and Williams, Ronald J},
  journal={nature},
  volume={323},
  number={6088},
  pages={533--536},
  year={1986},
  publisher={Nature Publishing Group}
}

@article{izmailov2018averaging,
  title={Averaging weights leads to wider optima and better generalization},
  author={Izmailov, Pavel and Podoprikhin, Dmitrii and Garipov, Timur and Vetrov, Dmitry and Wilson, Andrew Gordon},
  journal={arXiv preprint arXiv:1803.05407},
  year={2018}
}

@article{Czarnecki2020RealWG,
  title={Real World Games Look Like Spinning Tops},
  author={Wojciech M. Czarnecki and Gauthier Gidel and Brendan D. Tracey and Karl Tuyls and Shayegan Omidshafiei and David Balduzzi and Max Jaderberg},
  journal={ArXiv},
  year={2020},
  volume={abs/2004.09468},
  url={https://api.semanticscholar.org/CorpusID:215827540}
}

@article{vinyals2019grandmaster,
  title={Grandmaster level in StarCraft II using multi-agent reinforcement learning},
  author={Vinyals, Oriol and Babuschkin, Igor and Czarnecki, Wojciech M and Mathieu, Micha{\"e}l and Dudzik, Andrew and Chung, Junyoung and Choi, David H and Powell, Richard and Ewalds, Timo and Georgiev, Petko and others},
  journal={Nature},
  volume={575},
  number={7782},
  pages={350--354},
  year={2019},
  publisher={Nature Publishing Group}
}

@article{peng2017multiagent,
  title={Multiagent bidirectionally-coordinated nets: Emergence of human-level coordination in learning to play starcraft combat games},
  author={Peng, Peng and Wen, Ying and Yang, Yaodong and Yuan, Quan and Tang, Zhenkun and Long, Haitao and Wang, Jun},
  journal={arXiv preprint arXiv:1703.10069},
  year={2017}
}

@article{brown1951iterative,
  title={Iterative solution of games by fictitious play},
  author={Brown, George W},
  journal={Act. Anal. Prod Allocation},
  volume={13},
  number={1},
  pages={374},
  year={1951}
}

@article{omidshafiei2019alpha,
    author = {Shayegan Omidshafiei and Christos Papadimitriou and Georgios Piliouras and Karl Tuyls and Mark Rowland and Jean-Baptiste Lespiau and Wojciech M Czarnecki and Marc Lanctot and Julien Perolat and Remi Munos},
    title = {α-Rank: Multi-Agent Evaluation by Evolution},
    journal = {Scientific Reports},
    volume = {9},
    number = {1},
    pages = {1--29},
    year = {2019}
}

@inproceedings{liu2022unified,
  title={A Unified Diversity Measure for Multiagent Reinforcement Learning},
  author={Liu, Zongkai and Yu, Chao and Yang, Yaodong and Wu, Zifan and Li, Yuan and others},
  booktitle={Advances in Neural Information Processing Systems},
  year={2022}
}

@article{rame2024rewarded,
  title={Rewarded soups: towards pareto-optimal alignment by interpolating weights fine-tuned on diverse rewards},
  author={Rame, Alexandre and Couairon, Guillaume and Dancette, Corentin and Gaya, Jean-Baptiste and Shukor, Mustafa and Soulier, Laure and Cord, Matthieu},
  journal={Advances in Neural Information Processing Systems},
  volume={36},
  year={2024}
}

@article{munos2023nash,
  title={Nash learning from human feedback},
  author={Munos, R{\'e}mi and Valko, Michal and Calandriello, Daniele and Azar, Mohammad Gheshlaghi and Rowland, Mark and Guo, Zhaohan Daniel and Tang, Yunhao and Geist, Matthieu and Mesnard, Thomas and Michi, Andrea and others},
  journal={arXiv preprint arXiv:2312.00886},
  year={2023}
}

@article{song2023ensemble,
  title={Ensemble reinforcement learning: A survey},
  author={Song, Yanjie and Suganthan, Ponnuthurai Nagaratnam and Pedrycz, Witold and Ou, Junwei and He, Yongming and Chen, Yingwu and Wu, Yutong},
  journal={Applied Soft Computing},
  pages={110975},
  year={2023},
  publisher={Elsevier}
}

@article{wu2020deep,
  title={Deep ensemble reinforcement learning with multiple deep deterministic policy gradient algorithm},
  author={Wu, Junta and Li, Huiyun},
  journal={Mathematical Problems in Engineering},
  volume={2020},
  pages={1--12},
  year={2020},
  publisher={Hindawi}
}

@inproceedings{sheikh2022dns,
  title={DNS: Determinantal point process based neural network sampler for ensemble reinforcement learning},
  author={Sheikh, Hassam and Frisbee, Kizza and Phielipp, Mariano},
  booktitle={International Conference on Machine Learning},
  pages={19731--19746},
  year={2022},
  organization={PMLR}
}

@article{sharma2022deepevap,
  title={DeepEvap: Deep reinforcement learning based ensemble approach for estimating reference evapotranspiration},
  author={Sharma, Gaurav and Singh, Amit and Jain, Surbhi},
  journal={Applied Soft Computing},
  volume={125},
  pages={109113},
  year={2022}
}

@article{buckman2018sample,
  title={Sample-efficient reinforcement learning with stochastic ensemble value expansion},
  author={Buckman, Jacob and Hafner, Danijar and Tucker, George and Brevdo, Eugene and Lee, Honglak},
  journal={Advances in neural information processing systems},
  volume={31},
  year={2018}
}

@inproceedings{nikishin2018improving,
  title={Improving stability in deep reinforcement learning with weight averaging},
  author={Nikishin, Evgenii and Izmailov, Pavel and Athiwaratkun, Ben and Podoprikhin, Dmitrii and Garipov, Timur and Shvechikov, Pavel and Vetrov, Dmitry and Wilson, Andrew Gordon},
  booktitle={Uncertainty in artificial intelligence workshop on uncertainty in Deep learning},
  year={2018}
}

@article{southey2012bayes,
  title={Bayes' bluff: Opponent modelling in poker},
  author={Southey, Finnegan and Bowling, Michael P and Larson, Bryce and Piccione, Carmelo and Burch, Neil and Billings, Darse and Rayner, Chris},
  journal={arXiv preprint arXiv:1207.1411},
  year={2012}
}

@book{ferguson1991models,
  title={Models for the Game of Liar’s Dice},
  author={Ferguson, Christopher P and Ferguson, Thomas S},
  year={1991},
  publisher={Springer}
}

@inproceedings{balduzzi2019open,
  title={Open-ended learning in symmetric zero-sum games},
  author={Balduzzi, David and Garnelo, Marta and Bachrach, Yoram and Czarnecki, Wojciech and Perolat, Julien and Jaderberg, Max and Graepel, Thore},
  booktitle={International Conference on Machine Learning},
  pages={434--443},
  year={2019},
  organization={PMLR}
}

@article{hansen2004dynamic,
  title={Dynamic programming for partially observable stochastic games},
  author={Hansen, Eric A and Bernstein, Daniel S and Zilberstein, Shlomo},
  journal={Conference on Artificial Intelligence (AAAI)},
  year={2004}
}

@article{timbers2020approximate,
  title={Approximate exploitability: Learning a best response in large games},
  author={Timbers, Finbarr and Bard, Nolan and Lockhart, Edward and Lanctot, Marc and Schmid, Martin and Burch, Neil and Schrittwieser, Julian and Hubert, Thomas and Bowling, Michael},
  journal={arXiv preprint arXiv:2004.09677},
  year={2020}
}

@inproceedings{uchendu2023jump,
  title={Jump-start reinforcement learning},
  author={Uchendu, Ikechukwu and Xiao, Ted and Lu, Yao and Zhu, Banghua and Yan, Mengyuan and Simon, Jos{\'e}phine and Bennice, Matthew and Fu, Chuyuan and Ma, Cong and Jiao, Jiantao and others},
  booktitle={International Conference on Machine Learning},
  pages={34556--34583},
  year={2023},
  organization={PMLR}
}

@inproceedings{NEURIPS2021_fb4c4860,
 author = {Rebuffi, Sylvestre-Alvise and Gowal, Sven and Calian, Dan Andrei and Stimberg, Florian and Wiles, Olivia and Mann, Timothy A},
 booktitle = {Advances in Neural Information Processing Systems},
 editor = {M. Ranzato and A. Beygelzimer and Y. Dauphin and P.S. Liang and J. Wortman Vaughan},
 pages = {29935--29948},
 publisher = {Curran Associates, Inc.},
 title = {Data Augmentation Can Improve Robustness},
 url = {https://proceedings.neurips.cc/paper_files/paper/2021/file/fb4c48608ce8825b558ccf07169a3421-Paper.pdf},
 volume = {34},
 year = {2021}
}

@article{morales2024exponential,
  title={Exponential moving average of weights in deep learning: Dynamics and benefits},
  author={Morales-Brotons, Daniel and Vogels, Thijs and Hendrikx, Hadrien},
  journal={Transactions on Machine Learning Research},
  year={2024}
}

@article{he2019asymmetric,
  title={Asymmetric valleys: Beyond sharp and flat local minima},
  author={He, Haowei and Huang, Gao and Yuan, Yang},
  journal={Advances in neural information processing systems},
  volume={32},
  year={2019}
}

@article{liu2024neural,
  title={Neural Population Learning beyond Symmetric Zero-sum Games},
  author={Liu, Siqi and Marris, Luke and Lanctot, Marc and Piliouras, Georgios and Leibo, Joel Z and Heess, Nicolas},
  journal={arXiv preprint arXiv:2401.05133},
  year={2024}
}

@article{smith2021iterative,
  title={Iterative empirical game solving via single policy best response},
  author={Smith, Max Olan and Anthony, Thomas and Wellman, Michael P},
  journal={arXiv preprint arXiv:2106.01901},
  year={2021}
}

@article{li2024self,
  title={Self-adaptive PSRO: Towards an Automatic Population-based Game Solver},
  author={Li, Pengdeng and Li, Shuxin and Yang, Chang and Wang, Xinrun and Huang, Xiao and Chan, Hau and An, Bo},
  journal={arXiv preprint arXiv:2404.11144},
  year={2024}
}

@inproceedings{li2023solving,
  title={Solving large-scale pursuit-evasion games using pre-trained strategies},
  author={Li, Shuxin and Wang, Xinrun and Zhang, Youzhi and Xue, Wanqi and {\v{C}}ern{\`y}, Jakub and An, Bo},
  booktitle={Proceedings of the AAAI Conference on Artificial Intelligence},
  volume={37},
  pages={11586--11594},
  year={2023}
}

@article{li2024grasper,
  title={Grasper: A Generalist Pursuer for Pursuit-Evasion Problems},
  author={Li, Pengdeng and Li, Shuxin and Wang, Xinrun and Cerny, Jakub and Zhang, Youzhi and McAleer, Stephen and Chan, Hau and An, Bo},
  journal={arXiv preprint arXiv:2404.12626},
  year={2024}
}

@inproceedings{wang2023warm,
  title={Warm-start actor-critic: From approximation error to sub-optimality gap},
  author={Wang, Hang and Lin, Sen and Zhang, Junshan},
  booktitle={International Conference on Machine Learning},
  pages={35989--36019},
  year={2023},
  organization={PMLR}
}

\newpage
\appendix
\section{Algorithm for Fusion-PSD-PSRO}
\label{appendix-fusion-psd}
\begin{algorithm}
\caption{Fusion-PSD-PSRO}
\begin{algorithmic}[1] 
\State \textbf{Input:} initial policy sets for all players $\Pi$
\State Compute utilities $U^{\Pi}$ for each joint $\pi \in \Pi$
\State Initialize Meta-NE $\sigma_i = \text{UNIFORM}(\Pi_i)$
\For{\emph{e} $\in \{1, 2, \dots\}$}
    \For{\emph{player} $i \in \{1, 2, \dots, n\}$}
        \State Initialize $\pi^{t+1}_i$ via \textbf{Algorithm} \ref{Fusion} and sample $J$ policies $\{\pi^j_i\}^J_{j=1}$ from Policy Hull $\Pi^t_i$
        \For{many episodes}
            \State Sample $\pi_{-i} \sim \sigma^t_{-i}$ and collect the trajectory $\tau$ by playing $\pi_i$ against $\pi_{-i}$
            \State Discount the terminal reward $u(\pi_i, \pi_{-i})$ to each state as the extrinsic reward $r_1$
            \State Discount $R^{kl}(\tau)$ to each state as the intrinsic reward $r_2$
            \State Store $(s,a,s^{\prime},r)$ to the buffer, where $s^{\prime}$ is the next state and $r = r_1 + r_2$
            \State Estimate the gradient with the samples in the buffer and train oracle $\pi^{t+1}_i$ over $\rho \sim (\pi^{t+1}_i, \pi_{-i})$
        \EndFor
        \State $\Pi^{t+1}_{i} = \Pi^{t}_{i} \cup \{\pi^{t+1}_i\}$
    \EndFor
    \State Compute missing entries in $M^{t+1}$
    \State Compute a Meta-NE $\sigma$ from $M^{t+1}$
\EndFor
\State \textbf{Output:} current Meta-NE for each player.
\end{algorithmic}
\end{algorithm}

\section{Additional Experienments}
\begin{figure*}[tbp!]
    \centering
    \begin{subfigure}[b]{0.48\textwidth}
        \centering
        \includegraphics[width=\textwidth]{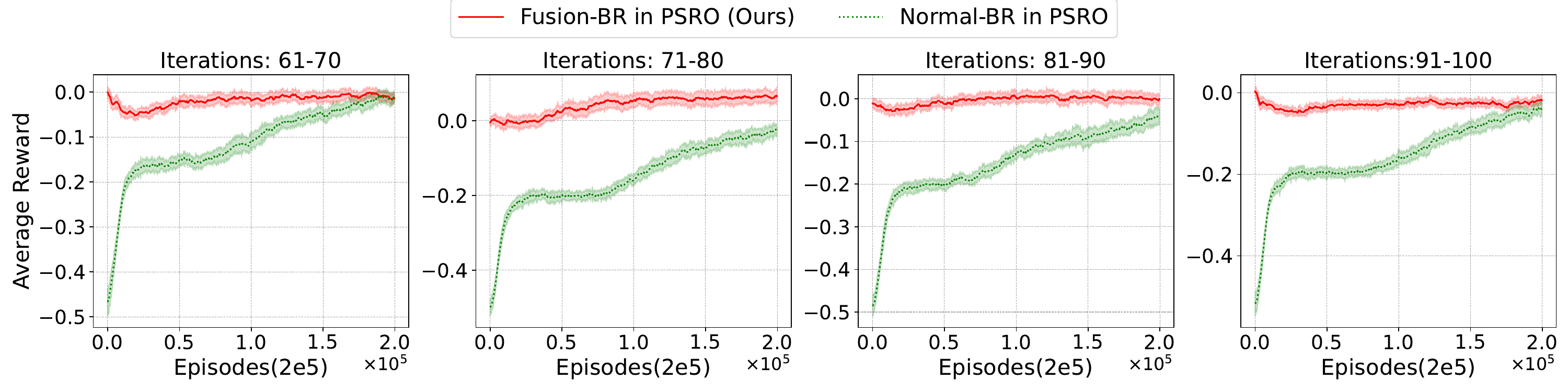}
    \end{subfigure}
    \begin{subfigure}[b]{0.48\textwidth}
        \centering
        \includegraphics[width=\textwidth]{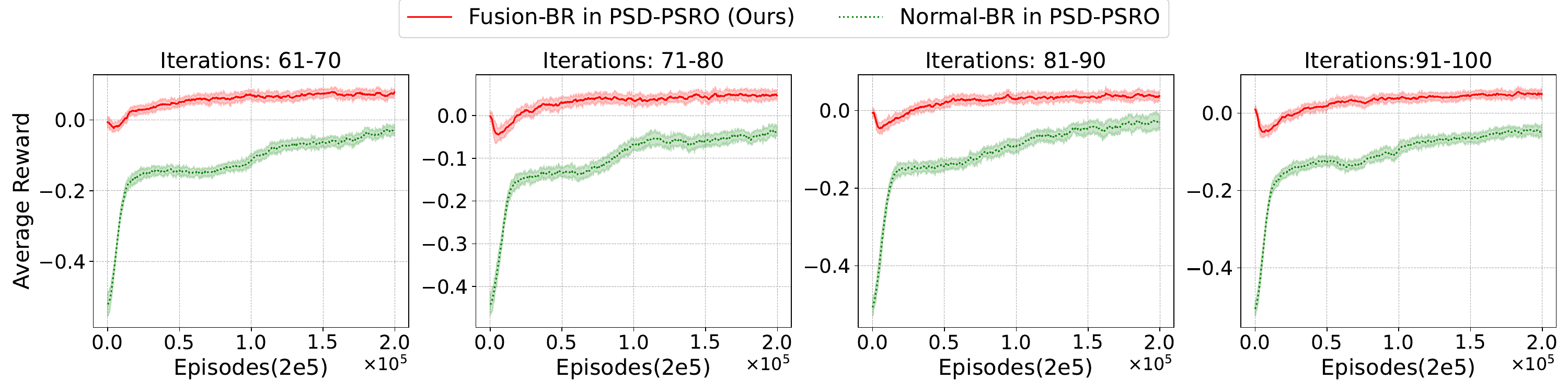}
    \end{subfigure}
    
    \begin{subfigure}[b]{0.48\textwidth}
        \centering
        \includegraphics[width=\textwidth]{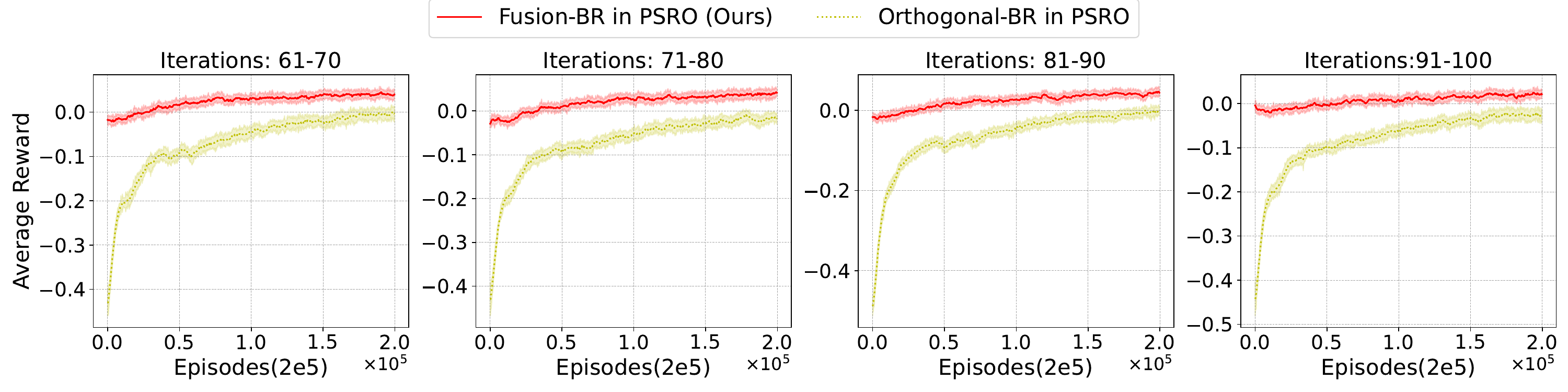}
    \end{subfigure}
    \begin{subfigure}[b]{0.48\textwidth}
        \centering
        \includegraphics[width=\textwidth]{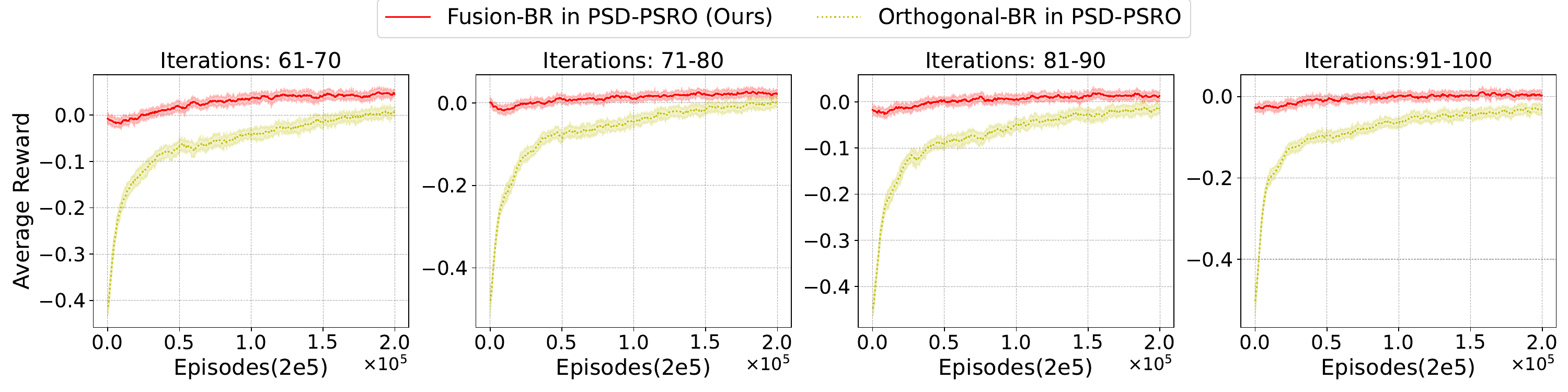}
    \end{subfigure}
    
    \begin{subfigure}[b]{0.48\textwidth}
        \centering
        \includegraphics[width=\textwidth]{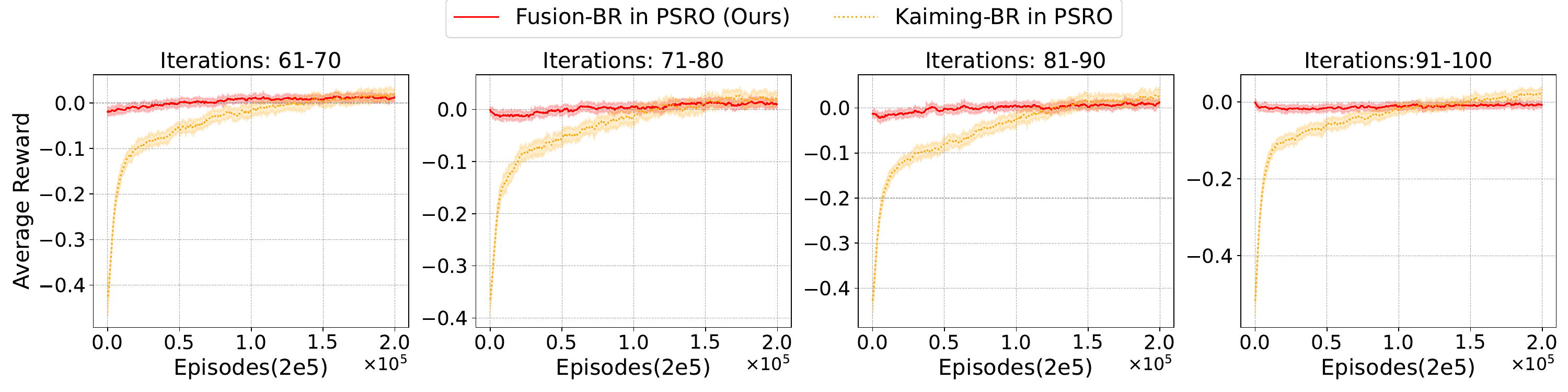}
    \end{subfigure}
    \begin{subfigure}[b]{0.48\textwidth}
        \centering
        \includegraphics[width=\textwidth]{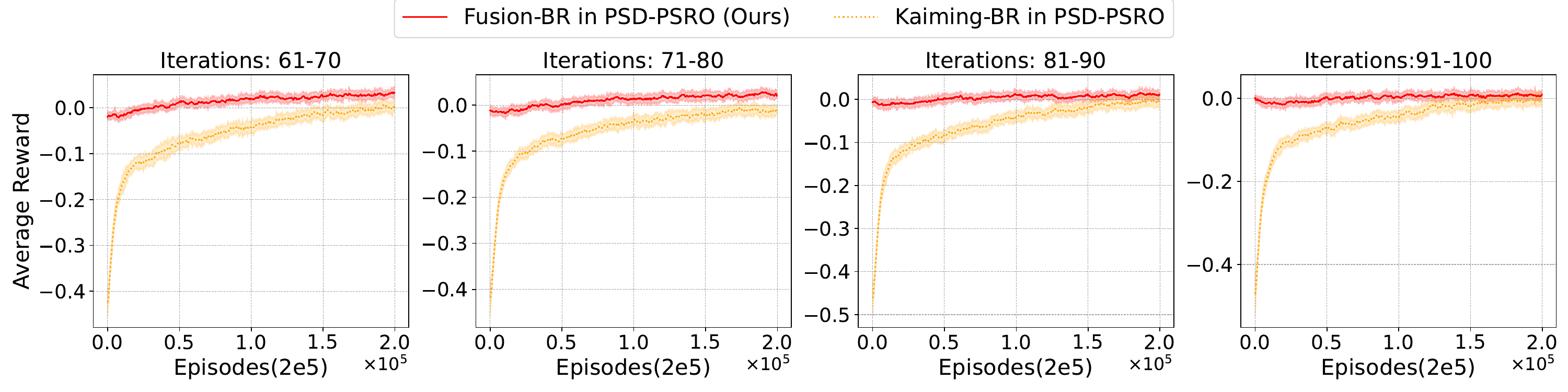}
    \end{subfigure}
    
    \begin{subfigure}[b]{0.48\textwidth}
        \centering
        \includegraphics[width=\textwidth]{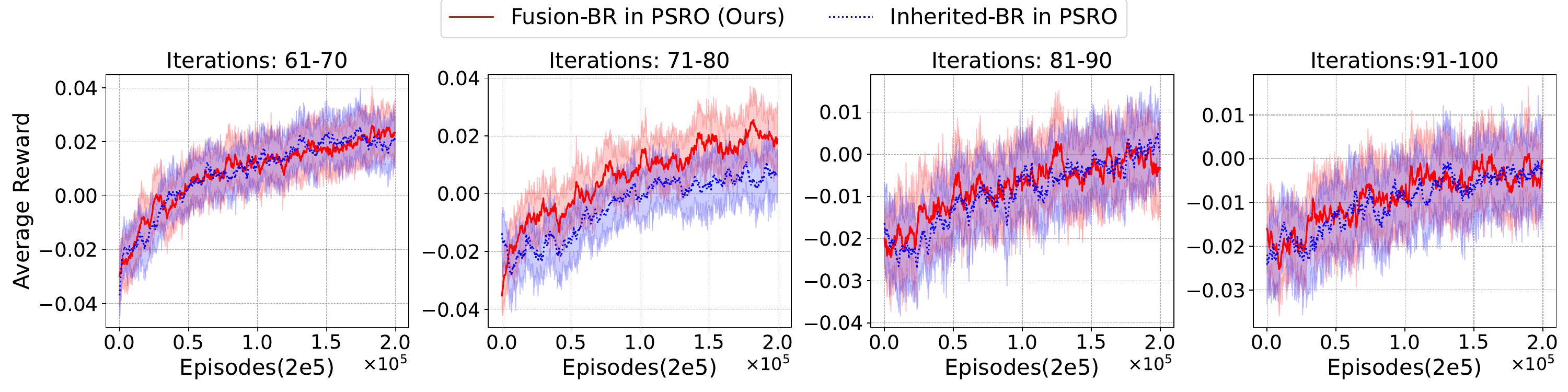}
    \end{subfigure}
    \begin{subfigure}[b]{0.48\textwidth}
        \centering
        \includegraphics[width=\textwidth]{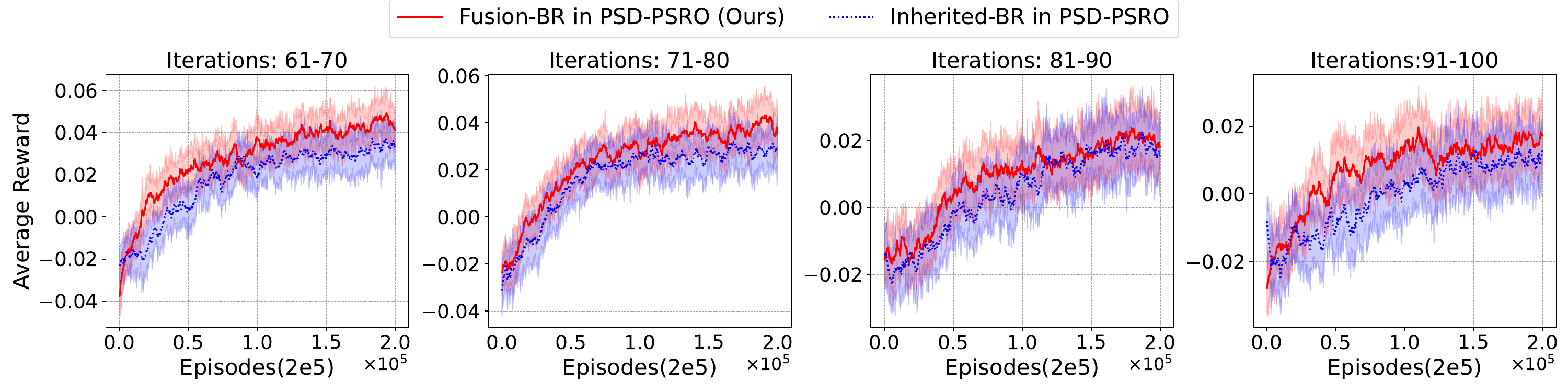}
    \end{subfigure}
    
    \caption{Average \emph{Reward} for each iteration (61-100) during the training of each approximate BR within PSRO and PSD-PSRO. Part 1 (top row) shows initial comparisons, while Part 2 (bottom rows) provides detailed comparisons across multiple reward settings.}
    \label{appendixreward}
\end{figure*}

\section{Average Reward during Training on Liar's Dice}
\label{appendix_reward}
In this ablation experiments, we initialized the initial populations using these four initialization methods respectively and continued using them to generate approximate BRs (corresponding to Normal-BR, Orthogonal-BR, Kaiming-BR and Inherited-BR respectively) during new policy training within PSRO and PSD-PSRO. Simultaneously, we replaced these initialization methods with the Nash Policy Fusion method to generate the second BR (Fusion-BR) for comparison. Fig.~\ref{appendixreward} shows the average \emph{reward} for each iteration (61-100) during the training of each approximate BR within PSRO and PSD-PSRO.

\newpage
\section{Benchmark and Implementation Details}
\label{appendix-benchmark}
\subsection{Non-Transitive Mixture Game}
This game is characterized by seven Gaussian humps that are evenly spaced on a two-dimensional plane. Each policy in the game corresponds to a point on this plane, analogous to the weights (the probability of that point within each Gaussian distribution) that players allocate to the humps. The payoff containing both non-transitive and transitive components is $\pi^T_i S \pi_{-i} + \frac{1}{2} \sum^7_{k=1} (\pi^k_i - \pi^k_{-i})$, where
\[
S = \begin{bmatrix}
0 & 1 & 1 & 1 & -1 & -1 & -1 \\
-1 & 0 & 1 & 1 & 1 & -1 & -1 \\
-1 & -1 & 0 & 1 & 1 & 1 & -1 \\
-1 & -1 & -1 & 0 & 1 & 1 & 1 \\
1 & -1 & -1 & -1 & 0 & 1 & 1 \\
1 & 1 & -1 & -1 & -1 & 0 & 1 \\
1 & 1 & 1 & -1 & -1 & -1 & 0 \\
\end{bmatrix}.
\]
Contrary to conventional approaches, where PSRO and PSRO$_{rN}$ are typically limited by the number of threads of only 1—compared to 4 in other PSRO variants—thus severely constraining their exploration capabilities, we opted for a different way. To mitigate this constraint, we tripled the training episodes for these two algorithms compared to others, aiming to balance the exploration limitations. 

\begin{table}[htbp!]
\caption{Hyperparameters for Leduc Poker.}
\centering
\small
\begin{tabular}{|l|l|}
 \hline
 \textbf{Hyperparameters} & \textbf{Value} \\
 \hline
 \multicolumn{2}{|l|}{\textit{Oracle}} \\
 \hline
 Oracle agent & DQN\\
 Replay buffer size & $10^4$ \\
 Mini-batch size & 512\\
 Optimizer & Adam \\
 Learning rate & $5 \times 10^{-3}$\\
 Discount factor ($\gamma$) & 1\\
 Epsilon-greedy Exploration ($\epsilon$)& 0.05\\
 Target network update frequency& 5\\
 Policy network & MLP (256-256-256)\\
 Activation function in MLP & ReLu \\
 \hline
 \multicolumn{2}{|l|}{\textit{PSRO}} \\
 \hline
 Episodes for each BR training & $2 \times 10^4$ \\
 meta-policy solver & Nash \\
 \hline
 \multicolumn{2}{|l|}{\textit{PSD-PSRO}} \\
 \hline
 Episodes for each BR training & $2 \times 10^4$ \\
 Meta-strategy solver & Nash \\
 Diversity weight ($\lambda$) & 1 \\
 \hline
 \multicolumn{2}{|l|}{\textit{Policy Fusion}} \\
 \hline
 Fusion start iteration & 2 \\
 Fusion method & Nash Policy Fusion \\
 \hline
\end{tabular}
\label{table-leduc}
\end{table}

\begin{table}[htbp!]
\caption{Hyperparameters for Liar's Dice.}
\centering
\small
\begin{tabular}{|l|l|}
 \hline
 \textbf{Hyperparameters} & \textbf{Value} \\
 \hline
 \multicolumn{2}{|l|}{\textit{Oracle}} \\
 \hline
 Oracle agent & Rainbow-DQN\\
 Replay buffer size & $10^5$\\
 Mini-batch size & 512\\
 Optimizer & Adam \\
 Learning rate & $5 \times 10^{-4}$\\
 Learning rate decay&linear decay\\
 Discount factor ($\gamma$) & 0.99\\
 Epsilon-greedy Exploration ($\epsilon$)& 0.05\\
 Target network update frequency& 5\\
  Network soft update ratio&0.005\\
 Prioritized Experience Replay parameter&0.6\\
  Important sampling parameter&0.4\\
 Gradient clip&10\\
 Policy network & MLP (256-256-128)\\
 Activation function in MLP & ReLu \\
 \hline
 \multicolumn{2}{|l|}{\textit{PSRO}} \\
 \hline
 Episodes for each BR training & $2 \times 10^5$\\
 meta-policy solver & Nash \\
 \hline
 \multicolumn{2}{|l|}{\textit{PSD-PSRO}} \\
 \hline
 Episodes for each BR training & $2 \times 10^5$\\
 Meta-strategy solver & Nash \\
 Diversity weight ($\lambda$) & 1 \\
 \hline
 \multicolumn{2}{|l|}{\textit{Policy Fusion}} \\
 \hline
 Fusion start iteration & 2 \\
 Fusion method & Nash Policy Fusion \\
 \hline
\end{tabular}
\label{table-liarsdice}
\end{table}

\begin{table}[htbp!]
\caption{Hyperparameters for Goofspiel (5 point cards).}
\centering
\small
\begin{tabular}{|l|l|}
 \hline
 \textbf{Hyperparameters} & \textbf{Value} \\
 \hline
 \multicolumn{2}{|l|}{\textit{Oracle}} \\
 \hline
 Oracle agent & DQN\\
 Replay buffer size & $10^4$ \\
 Mini-batch size & 512\\
 Optimizer & Adam \\
 Learning rate & $5 \times 10^{-3}$\\
 Discount factor ($\gamma$) & 1\\
 Epsilon-greedy Exploration ($\epsilon$)& 0.05\\
 Target network update frequency& 5\\
 Policy network & MLP (512-512-512)\\
 Activation function in MLP & ReLu \\
 \hline
 \multicolumn{2}{|l|}{\textit{PSRO}} \\
 \hline
 Episodes for each BR training & $3 \times 10^4$ \\
 meta-policy solver & Nash \\
 \hline
 \multicolumn{2}{|l|}{\textit{PSD-PSRO}} \\
 \hline
 Episodes for each BR training & $3 \times 10^4$ \\
 meta-strategy solver & Nash \\
 diversity weight ($\lambda$) & 1 \\
 \hline
 \multicolumn{2}{|l|}{\textit{Policy Fusion}} \\
 \hline
 Start fusion iteration & 20 \\
 Fusion method & Nash Policy Fusion \\
 \hline
\end{tabular}
\label{table-gf5}
\end{table}
\newpage
\subsection{Leduc Poker}
Since Diverse-PSRO cannot scale to the RL setting and the code for BD\&RD-PSRO in complex games is unavailable, we compare PSRO and PSD-PSRO with their policy fusion counterparts, Fusion-PSRO and Fusion-PSD-PSRO. We implement the PSRO paradigm with Meta-Nash solver, using DQN as the oracle agent. Hyper-parameters are shown in Table~\ref{table-leduc}. 

\subsection{Liar's Dice}
The Liar's Dice game involves two players, each equipped with a single die.  Incorrect challenges or bids result in immediate loss of the game, as they lead to the loss of dice. We implement the PSRO paradigm with Nash solver, using Rainbow-DQN as the oracle agent. Hyper-parameters are shown in Table \ref{table-liarsdice}, and the same set of parameters is used for Liar's Dice IR as well.

\subsection{Goofspiel}
In Goofspiel, we use DQN as the oracle agent. Hyper-parameters are shown in Table~\ref{table-gf5}. The start fusion iteration is set to 20.

\begin{figure}[!t]
    \centering
    \includegraphics[width=0.5\textwidth]{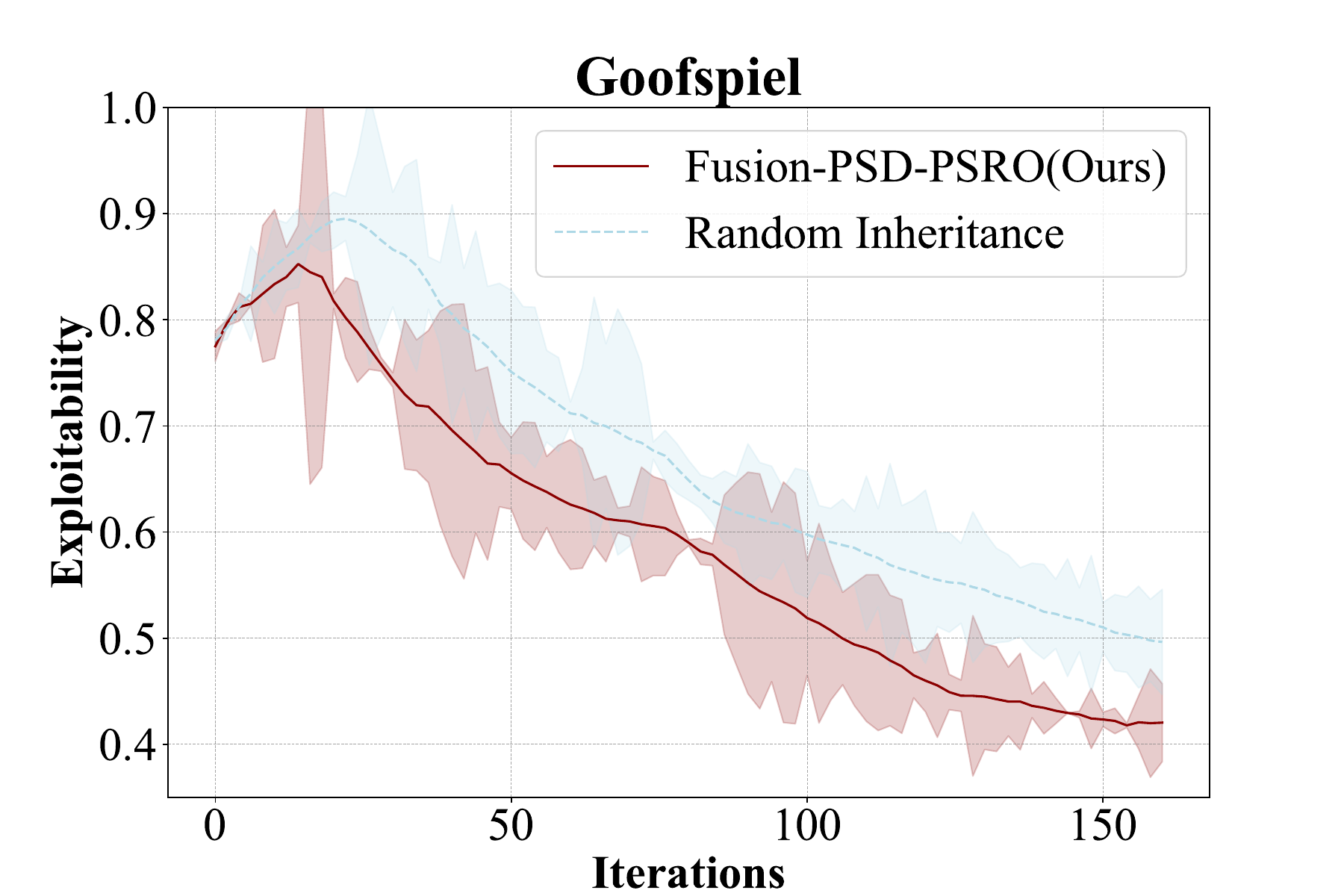}
    \caption{Exploitability Comparison: Fusion vs. Random Inheritance in Goofspiel (PSD-PSRO)}
    \label{gf5_compare_randon_inherit}
    \vspace{1.2\baselineskip}
\end{figure}

\subsection{Statistical Comparison with Random Inheritance}

The following analysis compares exploitability metrics between random inheritance baselines and our fusion methods in two imperfect-information games. Lower exploitability values indicate stronger strategy robustness.

\begin{table}[h!]
\centering
\caption{Exploitability Statistics for Liar's Dice (100 iterations)}
\label{tab:liars-dice-stats}
\begin{tabular}{lcccc}
\toprule
\textbf{Method} & \textbf{Mean} & \textbf{Max} & \textbf{Min} \\
\midrule
Random Inherit (PSRO) & 0.0474 & 0.0556 & 0.0393 \\
Fusion-PSRO & \textbf{0.0107} & \textbf{0.0175} & \textbf{0.0039} \\
\cmidrule(lr){1-4}
Random Inherit (PSD) & 0.0376 & 0.0411 & 0.0341 \\
Fusion-PSD-PSRO & \textbf{0.0148} & \textbf{0.0264} & \textbf{0.0045} \\
\bottomrule
\end{tabular}
\end{table}

Table \ref{tab:liars-dice-stats} shows consistent superiority of fusion methods in Liar's Dice. Both Fusion-PSRO and Fusion-PSD-PSRO achieve substantially lower exploitability across all metrics compared to their random inheritance counterparts, demonstrating enhanced strategy robustness.

\begin{table}[h!]
\centering
\caption{Exploitability Statistics for Goofspiel (160 iterations)}
\label{tab:goofspiel-stats}
\begin{tabular}{lcccc}
\toprule
\textbf{Method} & \textbf{Mean} & \textbf{Max} & \textbf{Min} \\
\midrule
Random Inherit (PSD-PSRO) & 0.4836 & 0.5332 & 0.4340 \\
Fusion-PSD-PSRO & \textbf{0.4004} & \textbf{0.4127} & \textbf{0.3881} \\
\bottomrule
\end{tabular}
\end{table}

As evidenced in Table \ref{tab:goofspiel-stats}, the fusion approach maintains its advantage in Goofspiel, with notably reduced exploitability values and tighter performance ranges compared to random inheritance. This pattern is further illustrated in Figure \ref{gf5_compare_randon_inherit}, showing the evolutionary trajectories of both methods.

\subsection{Time Consumption in Fusion-PSRO}
We analyze the time consumption specific to Nash Policy Fusion within the Fusion-PSRO framework. As shown in Table~\ref{tab:time_consumption}, the computation of the fusion policy parameters accounts for only 0.01\% of the total time during training on Leduc Poker. This indicates that Nash Policy Fusion is an efficient addition to the PSRO without significantly impacting overall runtime.

\begin{table}[ht!]
    \centering
    \begin{tabular}{lr}  
        \hline
        \textbf{Main Part of Fusion-PSRO} & \textbf{Percentage (\%)} \\
        \hline
        Compute NE of the meta game       & 0.19 \\
        Compute the Approximate BR        & 81.85 \\
        Compute the Fusion Policy Parameters & 0.01 \\
        Compute the Payoff Matrix         & 17.69 \\
        \hline
    \end{tabular}
    \caption{Time Consumption in Fusion-PSRO on Leduc Poker.}
    \label{tab:time_consumption}
\end{table}

\newpage

\end{document}